\documentclass[english,pre,twocolumn,nofootinbib,floatfix]{revtex4}

\usepackage[T1]{fontenc}
\usepackage[latin1]{inputenc}
\usepackage{amsmath}
\usepackage{amssymb}
\usepackage{babel}
\usepackage{epsfig}
\usepackage{subfigure}
\def\lB{\lambda_{\rm B}}
\def\d#1{\!{\rm d}#1\,}
\def\vec#1{\mathbf{#1}}

\def\kB{k_{\rm B}}
\def\Zdlvo{Z_{>}}

\def\Nabla\nabla
\def\Heff{H_{\rm eff}}

\def\fourier#1{#1_{\vec k}}
\def\mM{\mu\mathrm{M}}
\def\nm{\mathrm{nm}}
\def\np{n}

\def\cs{c_{\rm s}}
\def\PhiD{\Phi_{\rm D}}
\def\PiD{\Pi_{\rm D}}
\def\Piid{\Pi_{\rm id}}
\def\tr{\mathrm{tr}}
\def\trc{\tr_{\rm c}}
\def\Tr{\mathrm{Tr}}

\begin{document}

\title{Volume terms for charged colloids: a grand-canonical treatment}

\author{Bas Zoetekouw}
\email{B.Zoetekouw@phys.uu.nl}

\author{René van Roij}
\email{R.vanRoij@phys.uu.nl}

\affiliation{Institute for Theoretical Physics, Utrecht University,
Leuvenlaan 4, 3584CE Utrecht, the Netherlands}

\date{
    \today
%    (\SVNDate, \SVNTime;
%    revision:~r\SVNLastChangedRevision)
}

\begin{abstract}
We present a study of thermodynamic properties of suspensions of
charged colloids on the basis of linear Poisson--Boltzmann theory.
We calculate the effective Hamiltonian of the colloids by
integrating out the ionic degrees of freedom grand-canonically.
This procedure not only yields the well-known pairwise
screened-Coulomb interaction between the colloids, but also
additional volume terms which affect the phase behavior and the
thermodynamic properties such as the osmotic pressure. These
calculations are greatly facilitated by the grand-canonical
character of our treatment of the ions, and allow for relatively
fast computations compared to earlier studies in the canonical
ensemble. Moreover, the present derivation of the volume terms are
relatively simple, make a direct connection with Donnan
equilibrium, yield an explicit expression for the effective
screening constant, and allow for extensions to include, for
instance, nonlinear effects.
\end{abstract}
\maketitle

\section{Introduction}
\noindent Colloidal suspensions are multi-component systems that
consist of mesoscopic colloidal particles dispersed in a molecular
solvent. Often other chemical components are present as well,
e.g.\ ions, polymers, or proteins. Predicting or understanding the
properties of such mixtures from a microscopic perspective is
generally complicated, because the large asymmetry in size and charge
between the colloids and the other components in practice inhibits
a treatment of all the components on an equal footing. The
standard way out is to regard the suspension as an effective
colloids-only system, in which all microscopic degrees of freedom
of the ``medium'' (solvent, ions, polymers, etc.) are suitably
averaged out.

For instance, in the case of colloidal hard spheres in a medium
with non-adsorbing ideal polymers (radius of gyration $R_{\rm g}$)
the so-called depletion effect \cite{AO-depl,vrij-depl} leads to
effective attractions between pairs of colloids at surface-surface
separations less than $2R_{\rm g}$, and in the case of charged
colloidal spheres in an electrolyte with Debye length
$\kappa^{-1}$ the effective interaction between a colloidal pair
at center-to-center separation $r$ is generally written as a
repulsive screened-Coulomb potential $\propto\exp(-\kappa r)/r$.
The advantage of such a one-component viewpoint is that all of the
machinery of classical one-component fluids (integral equations,
perturbation theory, simulation, etc.) can be employed to study
the properties of colloidal suspensions, but only after a reliable
averaging over the medium has been performed. Performing this
averaging explicitly is generally a tremendous statistical
mechanics problem, that can only be solved approximately in most
cases~\cite{levin-review,hansen-lowen-review,likos-review}.

One important problem is that the effective colloidal interactions
are not necessarily pairwise additive, i.e.\ triplet or
higher-order many-body potentials may appear even if the
underlying interactions in and with the medium are strictly
pairwise. On physical grounds one generally expects the breakdown
of pairwise additivity of the effective interactions if the
typical length scale in the medium is of the order of the typical
colloidal length scale, e.g.\ the colloidal radius $a$. For
colloid--polymer mixtures it was indeed shown that equal-sized
colloids and polymers ($R_{\rm g}=a$) have bulk and interfacial
properties that differ dramatically from pairwise predictions
\cite{PRL-89-208303,JPCM-14-L1}, and charged colloids in an electrolyte were
shown to exhibit non-negligible effective triplet attractions on top of the
pairwise repulsions \cite{PRE-66-011402} at (extremely) low salt
concentrations where $\kappa^{-1}\sim a$.

\medskip

In this paper we will focus on a description of effective
interactions (or the effective Hamiltonian) in bulk suspensions of
charged colloids. The classical theory for these systems dates
back to the 1940's, when Derjaguin and Landau \cite{dl1941} and
Verweij and Overbeek \cite{vo1948} independently calculated that
the effective potential between {\em two} identical homogeneously
charged colloidal spheres (radius $a$, total charge $-Ze$ with $e$
the proton charge) in a bulk medium with dielectric constant
$\epsilon$ and Debye length $\kappa^{-1}$ is given by
\begin{equation}\label{V2gen}
V_2(r)=\left\{\begin{array}{ll}\displaystyle\infty & r<2a\\
\displaystyle\frac{Z^2e^2}{\epsilon}\left(\frac{\exp(\kappa
a)}{1+\kappa a}\right)^2\frac{\exp(-\kappa
r)}{r};&r>2a.\end{array}\right.
\end{equation}
Here and in the remainder of this paper we ignore the dispersion
forces, and we recall that $\kappa$ is defined as
\begin{equation}
\kappa^2=8\pi\lB {\cs} \label{kappa}
\end{equation}
in the case of a 1:1 electrolyte with total ion concentration
$2\cs$ far from the colloids, where $\lB=\beta e^2/\epsilon$
is the Bjerrum length, $\beta=1/\kB T$, and $T$ the 
temperature~\cite{dl1941,vo1948}.

It is well established by now that many properties of suspensions
of $N\gg1$ charged colloids in a solvent volume $V$ (density
$n=N/V$) can be understood on the basis of the pairwise effective
Hamiltonian
\begin{equation}
H_2(\{{\vec R}\})=\sum_{i<j}^NV_2(R_{ij}),\label{H2gen}
\end{equation}
where ${\vec R}_i$ denotes the position of colloid $i=1,\dots,N$,
and where $R_{ij}=|{\vec R}_i-{\vec R}_j|$. For instance, the
thermodynamic equilibrium properties and the phase behavior follow
from the Helmholtz free energy $F_2(N,V,T)$, defined as the
classical canonical phase space integral
\begin{align}\begin{split}
\exp(-\beta F_2)=&\frac{1}{N!{\cal V}^N}\int_V \d{\vec R}^N
\exp(-\beta H_2)\\
\equiv& \trc \exp(-\beta H_2),\label{F2}
\end{split}\end{align}
where ${\cal V}$ is an irrelevant constant volume (accounting for
the internal partition function of the colloids) that we include
for dimensional reasons, and where we introduced the short-hand
notation~$\trc$ for the classical canonical trace over the
colloid degrees of freedom. On the basis of Eqs.~\eqref{V2gen},
\eqref{H2gen}, and \eqref{F2} one can explain many experimental
observations, including the crystallization of (essentially) hard
spheres ($Z=0$ or $\kappa a\gg 1$) at packing fractions $\eta=4\pi
a^3n/3>0.5$ into an FCC crystal \cite{JCP-27-1208,PRL-63-2753},
the crystallization into BCC crystals for sufficiently soft
spheres \cite{EPL-12-81,PRL-94-138303}, the measured osmotic
equation of state \cite{reus1997,reus1999}, structure factor
\cite{PRL-62-1524,scatter1}, radial distribution function \cite{paddy2003},
pair interactions \cite{PRL-73-352,Langmuir-10-1351}, and many
other colloidal phenomena. It is therefore fair to state that the
DLVO theory as described by the Eqs.~\eqref{V2gen},
\eqref{H2gen}, and~\eqref{F2} is one of the corner stones of
colloid science.

It is, however, also fair to add that not all experimental
observations are in (qualitative) agreement with DLVO theory. For
instance, the experimental observation of ``voids'' and ``Swiss
cheese'' structures in otherwise homogeneous suspensions have been
interpreted as manifestations of gas--liquid coexistence
\cite{ise1994,PRL-78-2660}, and the small lattice spacing of
colloidal crystals compared to the one expected on the basis of
the known density~$n$ was interpreted as evidence for gas--crystal
coexistence \cite{ise1983}. These possibilities seemed to be
confirmed by direct observations of (meta-)stable gas--crystal
coexistence \cite{Nature-385-230}, and a macroscopic gas--liquid
meniscus \cite{tataarora} although these observations were
disputed by others \cite{palberg1994,arora_reply}.

Despite the ongoing debates due to a lack of experimental
consensus these experimental results, which were all performed at
low ionic strength with ${\cs}$ in the $\mM$-regime, triggered a
lot of theoretical activities to find the source of cohesive
energy that stabilizes the dense liquid or crystal phase in
coexistence with a much more dilute gas phase.  The dispersion
forces would be the first natural candidate to provide the
cohesion, but their nanometer range is generally considered to be
too small to dominate over the electrostatic repulsions with a
range of $\kappa^{-1}\sim 100~\nm$ at these low salt
concentrations.

It was for instance found that ion--ion
correlations, which are ignored in the derivation of $V_2(r)$, can
lead to attractive contributions to the pair potential. However, the
effect is small and too short-ranged for monovalent ions at room
temperature in water \cite{EPL-12-81}.

Another avenue of research
considered the possibility of the breakdown of pairwise additivity
due to non-negligible effective triplet and higher-order forces.
Within Poisson--Boltzmann theory the triplet potential was
calculated, and turned out to be attractive indeed
\cite{PRE-66-011402}, thereby suggesting that many-body
interactions could be the source of cohesive energy. Phase
diagrams based on repulsive pair interactions~\eqref{V2gen} and
the attractive triplet potential indeed showed coexistence of a
dilute gas with very dense crystal phases (as well as
crystal--crystal coexistence) \cite{JPCM-15-S3549,PRE-69-061407},
while experimental evidence for the breakdown of pairwise
additivity was obtained by an inverse Ornstein--Zernike analysis of
measured colloidal radial distribution functions
\cite{EPL-69-468,JPCM-15-S3509,PRL-91-115502}, as well as by direct
measurement\cite{PRL-92-078301,PRE-69-031402}. So although pairwise
additivity seems to be breaking down at low salinity, it is yet
questionable whether an approach based on the explicit calculation
of triplet and higher-order potentials, if feasible at all, is very
efficient, as convergence is probably slow in the regime of strong
triplet attractions: there is hardly any justification to ignore the
four-body potential when including the triplet potential changes the
phase diagram completely compared to the pairwise case. This notion
was made explicit by a recent simulation study of the primitive
model (charged colloids and explicit microions) that underlies the
effective one-component system of Ref.~\cite{PRE-69-061407}: the
gas--crystal coexistence that was found with included triplet
interactions disappeared again in the simulations of the
multi-component simulation \cite{JCP-123-244902}.

An alternative representation of non-pairwise interactions is
based on density-dependent pair-potentials. Roughly speaking, this
implies that the explicit coordinate-dependence of higher-body
potentials is smeared out to reduce to density dependence in the
pair interactions. In the case of charged colloids it seems
natural to modify the form of the screening constant, such that
not only the background (reservoir) salt concentration $2{\cs}$ but
also the finite concentration $Zn$ of the counterions and the
hard-core exclusion from the colloidal volume is taken into
account. For instance, one replaces $\kappa$ by
$\tilde{\kappa}=\sqrt{4\pi\lB(2{\cs}+Zn)}$,
$\sqrt{4\pi\lB(2{\cs}+Zn)/(1-\eta)}$ or similar expressions
\cite{EPL-12-81,JPCM-10-1219,JCP-112-4683,EPL-53-86,warren-condmat-2000,
PRE-62-3855,JCP-116-8588,Lang-17-4202,JPCM-15-S3467}
that reduce in the dilute limit $n\rightarrow 0$ to $\kappa$ as
given by Eq.~\eqref{kappa}. Often $\tilde{\kappa}(n)>\kappa$, and
one could interpret the resulting reduction of the pairwise
repulsions due to the more efficient screening at higher density
as an effective attractive many-body effect.

Interestingly, however, a careful analysis of the total free
energy of the suspension reveals that a density-dependent
screening constant affects not only the pair-interactions but also
one-body contributions, such as the free energy of each colloid
with its ``own'' diffuse cloud of counterions
\cite{JCIS-105-216,JCP-116-8588,Lang-17-4202,PRL-79-3082,JCP-112-4683,EPL-53-86,
warren-condmat-2000,PRE-59-2010,PRE-62-3855}. The thickness of
this double layer is typically $\tilde{\kappa}^{-1}$, and hence
its typical (free) energy is of the order of
$-(Ze)^2/\epsilon(a+\tilde{\kappa}^{-1})$, i.e.~the Coulomb energy
of two charges $\pm Ze$ at separation $a+\tilde{\kappa}^{-1}$.
This term lowers progressively with increasing $n$ and thus
provides cohesive energy, whereas it is an irrelevant constant
offset of the free energy if a constant $\kappa$ is taken instead
of $\tilde{\kappa}(n)$. It was shown that the density-dependence
of these so-called {\em volume terms} could drive a gas--liquid
spinodal instability at low salt concentrations
\cite{PRL-79-3082,warren-condmat-2000,JCP-112-4683,PRE-59-2010},
and could hence (qualitatively) explain some of the puzzling
experimental
observations.

There are several reasons, however, to revisit the theories of
e.g.\
Refs.\cite{PRL-79-3082,warren-condmat-2000,JCP-112-4683,PRE-59-2010}.
First of all, they are formulated in the canonical ensemble (fixed
ion concentrations), which not only obscures its close
relationship with the classical Donnan theory for colloidal
suspensions \cite{donnan1924,overbeek1956}, but also unnecessarily
complicates the numerical calculation of phase diagrams as we will
argue in section~II.

Moreover, and more importantly, the derivation of the explicit
expressions for the total free energy was perhaps not very
transparent in Refs.~\cite{PRL-79-3082,PRE-59-2010}, and may have
hindered extensions of the theory to include, for instance, charge
renormalization. This nonlinear effect was first studied in a cell
geometry \cite{JCP-80-5776}, and, more recently, in a jellium-like
model\cite{PRE-69-031403,JPCM-15-S3523}.  In both of those models,
the nonlinear character of the theory is retained, while its
complicated multi-centered nature is replaced by a radially
symmetric structure.  The effective
colloidal charge $Z^*$ that appears in the prefactor of the DLVO
repulsions, is then reduced from its bare value $Z$ due to a tightly
adsorbed layer of counterions in the vicinity of the colloidal
surface. This effect is important when $Z\lB/a\gg 1$
\cite{JCP-80-5776,Lang-19-4027,CollSutfA-140-227,JCP-117-8138,EPL-41-123},
and therefore casts serious doubt\cite{JCP-119-1855} on the
predictions of the gas--liquid and gas--crystal transitions in e.g.
Refs.~\cite{PRL-79-3082,JCP-112-4683,PRE-59-2010} since large values
of $Z$ were needed to have the
transitions~\cite{JPCM-15-S3523,EPL-55-580}.  If one now realizes
that $Z^*$ depends on $n$ and $\tilde{\kappa}(n)$, as was shown in
e.g.\ Ref.~\cite{Lang-19-4027}, it is easy to imagine that the
volume terms are affected non-trivially by charge renormalization
similarly as by the $n$-dependence of the screening parameter. It is
therefore important to be able to include this effect into
volume-term-type theories, and hence to reformulate these theories
as transparently as possible.

\medskip

In order to be able to address all these issues, we revisit here
the purely linear screening theory with volume terms. Its
nonlinear extension to include charge renormalization will be
discussed in a forthcoming paper \cite{bas_multcel}. The present paper is
organized as follows. In section II we introduce the microscopic
Hamiltonian ${\cal H}$ of the colloid-ion mixture and give formal
expressions for the effective Hamiltonian $H$ for the colloids. In
section~III we calculate $H$ by minimizing the mean-field grand
potential functional of the ions, whereby explicit expression for
the density-dependent screening parameter, the Donnan potential,
and the Donnan effect are obtained as intermediate results. In
section~IV we consider the thermodynamics of the suspension, in
particular the free energy and the osmotic pressure, with a few
interesting canceling contributions. In section~V we calculate a
few phase diagrams. We conclude and summarize in section~VI.

\section{Hamiltonian, Donnan ensemble, and effective Hamiltonian}
\noindent
We consider a suspension of $N$~identical colloidal spheres
(radius~$a$, positions~${\vec R}_i$, charge~$-Ze$ homogeneously
distributed on the surface) in a continuum solvent of volume~$V$
characterized by a dielectric constant~$\epsilon$ at
temperature~$T$.  The density of the colloids is denoted
by~$n=N/V$.  In addition there are~$N_+$ and $N_-$~monovalent
point-like cations~($+$) and anions~($-$) present, respectively, and
charge neutrality dictates that $N_+=N_-+ZN$. The total interaction
Hamiltonian of the system can therefore be written as
\begin{equation}
{\cal H}={\cal H}_{\rm cc}+{\cal H}_{\rm cs}+{\cal H}_{\rm ss},\label{Hparts}
\end{equation}
where the bare colloid--colloid Hamiltonian ${\cal H}_{\rm cc}$, the
colloid--salt Hamiltonian ${\cal H}_{\rm cs}$, and the salt--salt
Hamiltonian ${\cal H}_{\rm ss}$ are pairwise sums of hard-core and
(unscreened) Coulomb potentials. We write ${\cal
H}_{\rm cc}=\sum_{i<j}^N V_{\rm cc}(R_{ij})$ with
\begin{equation}
\label{Vcc}
\beta V_{\rm cc}(r)= \begin{cases}
  \displaystyle
  \infty & r<2a;\\
  \displaystyle
  \frac{Z^2\lB}{r} \qquad &r>2a,
\end{cases}
\end{equation}
and ${\cal H}_{\rm cs}={\cal H}_{\rm c+}+{\cal H}_{\rm c-}$ with ${\cal
H}_{\rm c\pm}=\sum_{i=1}^N\sum_{j=1}^{N_\pm}V_{\rm c\pm}(|\vec
R_i-\vec r^\pm_j|)$, where
\begin{equation}
\label{Vcpm}
\beta V_{c\pm}(r)=\begin{cases}
  \displaystyle
  \infty & r<a;\\
  \displaystyle
  \mp\frac{Z\lB}{r} \qquad &r>a,
\end{cases}
\end{equation}
and $\vec r^\pm_j$ is the position of the $j$th positive (negative)
micro-ion.
The expression for ${\cal H}_{\rm ss}$ is similar, but without the
hard-core terms because of the point like nature of the ions.

In principle, the thermodynamic properties of this system could be
calculated from the Helmholtz free energy of the system
$\mathcal{F}(N,N_-,V,T)$, which is defined by
$\exp(-\beta\mathcal{F})=\trc \tr_+ \tr_-\exp(-\beta
{\cal H})$. The canonical traces are defined as in
Eq.~\eqref{F2}. Note that one can ignore the explicit $N_+$
dependence of~${\cal F}$ because of the charge neutrality
condition.

Within linearized Poisson--Boltzmann theory and
exploiting the Gibbs--Bogolyubov inequality, ${\cal F}$ was
explicitly calculated in Refs.~\cite{PRL-79-3082,PRE-59-2010}.
The phase diagram was then constructed from~$\mathcal{F}$ by
imposing the usual conditions of mechanical and diffusive
equilibrium, {\em viz.}
\begin{align}\label{cancoex}\begin{cases}
P(n^{(1)},n_-^{(1)})&=P(n^{(2)},n_-^{(2)})\\
\mu(n^{(1)},n_-^{(1)})&=\mu(n^{(2)},n_-^{(2)})\\
\mu_-(n^{(1)},n_-^{(1)})&=\mu_-(n^{(2)},n_-^{(2)}),
\end{cases}\end{align}
where $n^{(i)}$ and $n_-^{(i)}$ denote the colloid density and the
anion density in phase $i$, respectively, and where we introduced
the pressure $P=-(\partial{\cal F}/\partial V)$, the colloidal
chemical potential $\mu=(\partial{\cal F}/\partial N)$, and the
anion chemical potential $\mu_-=(\partial{\cal F}/\partial N_-)$.
The system \eqref{cancoex} of three equations for the four unknown
densities yielded the phase diagram in the $n-n_-$-plane, for
given parameters $Z$, $a$, and~$\lB$.

These canonical
ensemble calculations were, however, numerically rather demanding,
since many numerically expensive evaluations of $P(n,n_-)$,
$\mu(n,n_-)$ and $\mu_-(n,n_-)$ are needed in the root-finding
procedure of solving the Eqs.~\eqref{cancoex}. For that reason the
phase diagram of only a few combinations of parameters $Z$,
$\lB$, and $a$ has been studied in some detail. Moreover,
the derivation of the explicit expressions for ${\cal F}$ was
perhaps not very transparent, and may have hindered extensions of
the theory to include, for instance, nonlinear effects such as
charge renormalization. And on top of this the (strong) connection
with the standard description of colloidal suspensions in terms of
a Donnan equilibrium was not made in
Refs.~\cite{PRL-79-3082,PRE-59-2010}.

It turns out, as we will show in this paper, that at least some of
these shortcomings and drawbacks of working in the canonical
ensemble can be lifted by treating the anions and cations
grand-canonically. For this we assume the suspension to be in
diffusive contact with a dilute reservoir of monovalent anions and
cations at chemical potential $\mu_\pm =
\kB T\ln({\cs}\Lambda_\pm^3)$, where $2{\cs}$ is the total ion density
in the (charge neutral) reservoir, and where $\Lambda_\pm$ is the
thermal wavelength of the cations ($+$) and anions ($-$),
respectively. The colloidal particles cannot enter the ion reservoir
(e.g.\ because of a semi-permeable membrane in an actual
experimental setting), and remain treated canonically (fixed $N$ and
$V$) as before. The thermodynamic potential of this ensemble, which
we will call the ``Donnan-ensemble'' from now on, is denoted $F={\cal
F}-\mu_+N_+-\mu_-N_-$, and is a function of the variables $N$, $V$,
$T$, and $\mu_\pm$. It is related to the microscopic Hamiltonian by
the ``Donnan partition function''
\begin{equation}
\exp(-\beta F)=\trc\, \Tr_+\, \Tr_-\, \exp(-\beta {\cal
H}), \label{F}
\end{equation}
where ${\cal H}$ was defined in Eq.~\eqref{Hparts} and the grand canonical
traces are defined as
\begin{equation}
\Tr{}_\pm= \sum_{i=1}^{N_\pm}\exp(\beta\mu_\pm
N_\pm)\tr{}_\pm=
\sum_{i=1}^{N_\pm}\frac{{\cs}^{N_\pm}}{N_\pm!}\int \d{\vec
r^{N_\pm}_\pm}. \label{Tr}
\end{equation}
Here we have used that $\exp(\beta\mu_\pm)/\Lambda_\pm^3={\cs}$ (where
the factor $1/\Lambda_\pm^{3}$ follows from the classical momentum
integration), and we denoted the microion coordinates by ${\vec
r}^{N_\pm}_\pm$. For convenience we will drop the explicit
$T$-dependence from now on, and we replace the dependence on
$\mu_{\pm}$ by  the reservoir concentration ${\cs}$.

Because of the extensive character of $F$ we can write
$F(N,V,{\cs})=Vf(n,{\cs})$.  The thermodynamic properties follow now
as
$\mu=(\partial F/\partial N)=(\partial f/\partial \np)$
and
$P=-(\partial F/\partial V)=\np\mu-f$, where the derivatives are
to be taken at fixed ${\cs}$ and~$T$. This implies that the
phase-coexistence conditions simplify to the {\em two} conditions
\begin{align}\label{grcancoex}\begin{cases}
P(n^{(1)},{\cs})   &= P(n^{(2)},{\cs}); \\
\mu(n^{(1)},{\cs}) &= \mu(n^{(2)},{\cs}),
\end{cases}\end{align}
for the {\em two} unknown colloid densities $n^{(i)}$, at fixed
${\cs}$, i.e.\ we have prearranged equal chemical potential of the
ions due to our choice to work in the Donnan ensemble. This is a
considerable reduction of the numerical effort compared to the
Eqs.~\eqref{cancoex}. Note that the mechanical equilibrium
condition is equivalent to equal osmotic pressure $\Pi$ in the two
coexisting phases, where $\Pi(n,{\cs})=P(n,{\cs})-P(0,{\cs})$ is the
suspension's excess pressure over the reservoir pressure
$P(0,{\cs})=2\cs\kB T$ (recall that we treated the reservoir as an
ideal gas here). This simple relation allows for a rather direct
contact with experimental measurements of the osmotic pressure, as
we will also show below .

Even though our main objective is to calculate $F(N,V,{\cs})$ as
defined in Eq.~\eqref{F}, we will first focus on an important and
convenient intermediate quantity: the effective Hamiltonian $H$,
which depends on the colloid configuration $\{{\vec R}\}$ and
parametrically on the reservoir salt concentration~${\cs}$.
It is defined as
\begin{align}\begin{split}\label{ebH}
\exp&(-\beta H)
  = \Tr_+\,\Tr_-\,\exp(-\beta {\cal H}) \\
 &= \exp(-\beta {\cal H}_{\rm cc})\Tr_+\,\Tr_-\,\exp(-\beta
    {\cal H}_{\rm cs}-\beta{\cal H}_{\rm ss}) \\
 &\equiv \exp(-\beta {\cal H}_{\rm cc})\exp(-\beta\Omega)
\end{split}\end{align}
where, in the last step, we defined the grand partition function
$\exp(-\beta\Omega)$ of the {\em inhomogeneous} system of
interacting cations and anions (through ${\cal H}_{\rm ss}$) in the
external potential of the colloids (through ${\cal H}_{\rm cs})$.
The corresponding grand potential of this system is $\Omega$, which
is the quantity that we need to calculate in order to find the
effective Hamiltonian given from Eq.~\eqref{ebH} as
\begin{equation} \label{H}
H={\cal H}_{cc}+\Omega.
\end{equation}
Once $H$ is known, we can use standard one-component techniques to
obtain approximate expressions for $F$, since $\exp(-\beta
F)=\trc\,\exp(-\beta H)$ is precisely as if $F$ were the
Helmholtz free energy of a one-component system with Hamiltonian
$H$.

\section{The grand potential $\Omega$}
\subsection{Density functional}
\noindent
We will {\em not}
explicitly calculate $\Omega$ from the grand partition function of
Eq.~\eqref{ebH}. Instead we exploit the framework of classical density
functional theory (DFT), which treats an inhomogeneous fluid in an
external field at the level of the one-body distribution functions
(the density profiles) \cite{AdvPhys-28-143,Evans1992,JPCM-14-11897}: the
equilibrium density profiles minimize the (variational) grand
potential functional, and this minimum is the grand potential. Here
we denote the density profile of the cations by $\rho_+({\vec r})$,
that of the anions by $\rho_-({\vec r})$, and the grand-potential
functional by $\Omega[\rho_+,\rho_-]$. For notational convenience we
do not introduce a separate notation for the variational and
equilibrium profiles, and neither for the grand-potential
functional and its minimum (equilibrium) value.

The cations and anions experience external potentials $U_+({\vec
r})$ and $U_-({\vec r})$, respectively, due to the Coulomb and
excluded volume interactions with a fixed configuration $\{{\vec
R}\}$ of colloidal particles. These potentials are explicitly given
by
\begin{equation}
U_\pm({\vec r})=\sum_{i=1}^N V_{c\pm}(|{\vec R}_i-{\vec r}|)
\end{equation}
where the colloid--ion pair potentials $V_{\rm c\pm}(r)$ were
defined in Eq.~\eqref{Vcpm}.  We can now write the grand-potential
functional within a simple mean-field approximation as
\begin{eqnarray}
\Omega[\rho_+,\rho_-]&=&\Omega_{\rm id}[\rho_+]+ \Omega_{\rm id}[\rho_-]
%\nonumber\\
%&+&
+\frac{e^2}{2\epsilon}\int \d{\vec r}\d{\vec r'} \frac{\rho({\vec
r})\rho({\vec r}')}{|{\vec r}-{\vec r}'|}\nonumber \\
&+&\int \d{\vec r}\Big(\rho_+({\vec r})U_+({\vec r})+\rho_-({\vec
r})U_-({\vec r})\Big), \label{eq:functional}
\end{eqnarray}
where we defined the ion charge density $\rho({\vec r})=\rho_+({\bf
r})-\rho_-({\vec r})$, and where the ideal-gas grand potential
functional can be written as
\begin{eqnarray}
\Omega_{\rm id}[\rho_\pm]&=&\int \d{\vec r}\rho_\pm({\vec r})
\Big(-\mu_\pm+\kB T\big(\ln\rho_\pm({\vec
r})\Lambda_\pm^3-1\big)\Big)\nonumber\\
&=&\kB T\int \d{\vec r}\rho_\pm({\vec r})\Big(\ln\frac{\rho_{\pm}({\vec
r})}{{\cs}}-1\Big).
\end{eqnarray}
Here we have substituted the identity $\mu_\pm=\kB T\ln {\cs}\Lambda_\pm^3$.

The Euler--Lagrange equations $\delta\Omega/\delta\rho_\pm({\vec r})=0$ that
correspond with Eq.~\eqref{eq:functional}, can be cast, for ${\vec r}$ outside a
colloidal hard core, into the form $\rho_\pm({\vec r})={\cs}\exp[\mp\phi({\vec
r})]$.  The dimensionless potential $\phi({\vec r})$ must then
satisfy the nonlinear multi-centered Poisson--Boltzmann
equation~\cite{marcus1955}
\begin{equation}
\nabla^2\phi({\vec r})=\kappa^2\sinh\phi({\vec
r})-\frac{Z\lB}{a^2}\sum_{i=1}^N\delta(|{\vec r}-{\vec
R}_i|-a)\label{NPB},
\end{equation}
where $\delta(r)$ is the Dirac-delta. Unfortunately, no analytical
solution to Eq.~\eqref{NPB} is known for the multi-centered
geometry of interest here. Even solving Eq.~\eqref{NPB} numerically
is far from trivial, and requires a serious computational effort
\cite{fushiki_JCP_97,PRL-68-1081,JCP-92-3275,dobnikar2,dobnikar1,dobnikar3}.

For this reason we will first make further approximations to the
functional, and then perform its minimization afterwards. The main
approximation involves the expansion, up to quadratic order, of
the ideal-gas grand potential terms about the, as of yet unknown, ion
densities $\bar{\rho}_\pm$, such that $\rho_\pm(\vec{r})-\bar{\rho}_\pm$
are considered to  be the ``small'' expansion
parameters. This expansion yields
$\Omega_{\rm id}[\rho_{\pm}]\approx\Omega'_{\rm id}[\rho_\pm]$ with
\begin{eqnarray}
\beta\Omega'_{\rm id}[\rho_\pm]&=&
    \bar{\rho}_\pm \big(\ln\frac{\bar\rho_\pm}{{\cs}} - 1\big)V
    +\ln\frac{\bar\rho_\pm}{{\cs}}\int \d{\vec r}(\rho_{\pm}(\vec{r}) -
    \bar{\rho}_{\pm})\nonumber\\
    &+&\frac{1}{2\bar\rho_\pm}\int \d{\vec r}\big(\rho_\pm(\vec r) -
    \bar\rho_\pm\big)^2.
\label{Oidapprox}
\end{eqnarray}
In principle, this expansion holds for arbitrary $\bar{\rho}_\pm$,
but later on we will choose $\bar{\rho}_\pm$ to be equal to the
average ion concentrations in the system, such that $\int \d{\vec
r}(\rho_{\pm}({\vec r})-\bar{\rho}_{\pm})=0$,
i.e.\ $V\bar{\rho}_\pm=N_\pm$ is the number of ions in the suspension.
As will be shown below, this linearization corresponds to a
linearization of Eq.~\eqref{NPB} about $\phi({\vec r})=\bar{\phi}$
with $\bar{\phi}$ the Donnan potential. This is in line with
Ref.~\cite{PRE-66-011401}.

It turns out to be convenient, and necessary, to rewrite the
external potentials $U_\pm({\vec r})$ for the ions such  that
$U_\pm({\vec r})=\pm V({\vec r}) + W({\vec r})$, where we
defined the electrostatic potential (due to the colloids) $V({\vec
r})=\sum_i v(|{\vec r}-{\vec R}_i|)$ and the hard-core potential
$W({\vec r})=\sum_i w(|{\vec r}-{\vec R}_i|)$, with
\begin{subequations}
\begin{align}
\beta v(r) &= \begin{cases}
  \beta v_0   &  r<a; \\
  -Z\lB/r     &  r>a, \label{v}
\end{cases}
\intertext{and}
\beta w(r) &= \begin{cases}
  \beta w_0   &  r<a; \\
  0           &  r>a. \label{w}
\end{cases}
\end{align}
\end{subequations}
Although we are actually interested in the hard-core limit $\beta
v_0\rightarrow\infty$ and~$\beta
w_0\rightarrow\infty$ here, we introduce the (finite) hard-core
parameters $v_0$ and $w_0$ here for later convenience. They are
necessary and sufficient to ensure, within the linearized theory, a vanishing
ion density in the colloidal hard cores.

Collecting the results we can write the approximate
grand-potential functional as
\begin{align}
\Omega&[\rho_+,\rho_-] =
 \Omega'_{\rm id}[\rho_+]
 +\Omega'_{\rm id}[\rho_-]
 +\frac{e^2}{2\epsilon}\int \d{\vec r} \d{\vec r'}
     \frac{\rho({\vec r})\rho({\vec r}')}{|{\vec r}-{\vec r}'|} \nonumber \\
&+ \int \d{\vec r} \Big\{
  \rho({\vec r})V({\vec r}) + \big(
    \rho_+({\vec r})+\rho_-({\vec r})
  \big) W({\vec r})
\Big\},
\label{eq:functional2}
\end{align}
which is minimized by those (equilibrium) profiles that satisfy
the Euler--Lagrange equations
\begin{eqnarray}
\ln\frac{\bar\rho_\pm}{{\cs}}+\frac{\rho_\pm(\vec
r)-\bar\rho_\pm}{\bar\rho_\pm}
\pm \phi(\vec r) %\\&
+ \beta W(\vec r)=0. \label{el1}
\end{eqnarray}
Here we introduced the (dimensionless) electrostatic
potential~$\phi(\vec r)$, given by
\begin{align}\label{eq:phi}
\phi(\vec r) = \lB\int\d{\vec r'} \frac{\rho(\vec r')}{|\vec r -
\vec r'|} + \beta V(\vec r).
\end{align}

\subsection{Equilibrium profiles and Donnan equilibrium}
\noindent
We leave the hard-core parameters $v_0$ and $w_0$ undetermined for now,  and
start the analysis of the Euler--Lagrange equations by integrating
Eq.~\eqref{el1} over the volume.  At the same time, we impose that $\int
\d{\vec r}[\rho_\pm({\vec r})-\bar{\rho}_\pm]=0$, i.e.\ we choose
$\bar{\rho}_\pm$ such that it is the actual average ion density in the
suspension. After rearrangement, we find that
\begin{gather}
\label{elint} {\bar\rho_\pm} = {\cs}\exp[\mp \bar\phi - \eta\beta w_0],
\end{gather}
where $\bar\phi=\int\d{\vec r}\phi(\vec r)/V$ is the spatially
averaged electric potential, i.e.\ the Donnan potential.
Since global charge neutrality imposes that $\bar{\rho}_+-\bar{\rho}_-=Zn$,
we can conclude from Eq.~\eqref{elint} that
the Donnan potential satisfies
\begin{equation}
\label{donnan} \sinh\bar{\phi} =
  -\frac{Zn}{2{\cs}}\exp[\eta\beta w_0],
\end{equation}
which reduces to the usual Donnan expressions in the
point-colloid limit $\eta\rightarrow 0$
\cite{donnan1924,deserno-holm-2001}. Combining
Eq.~\eqref{donnan} with~\eqref{elint} yields
\begin{align}\label{eq:salt}
\bar\rho_\pm = \tfrac{1}{2}\left(
\sqrt{(Zn)^2+(2\cs)^2 \exp(-2\eta\beta w_0)} \pm Zn \right),
\end{align}
which explicitly relates the salt concentration in the suspension
to the colloid density and the salt reservoir concentration,
provided the parameter $w_0$ is known.

Using these relations for $\bar{\phi}$ and $\bar{\rho}_\pm$ we
consider a specific linear combination of the Euler--Lagrange
equations, and rewrite Eqs.~\eqref{el1} as
\begin{subequations}
\begin{align}
\label{el2a}
  \frac{\rho_+(\vec r)}{\bar\rho_+}
  + \frac{\rho_-(\vec r)}{\bar\rho_-} -2
  =& -2\left(\beta W(\vec r)-\eta\beta w_0\right); \\
  \begin{split}
    \rho(\vec r) - \bar\rho = &
    -(\bar\rho_++\bar\rho_-)(\phi(\vec r)-\bar\phi) \\
    &-\bar\rho \left(\beta W(\vec r) -\eta\beta w_0\right),
  \end{split}
\label{el2b}
\end{align}
\end{subequations}
where we defined the short-hand notation
$\bar{\rho}=\bar\rho_+-\bar\rho_-=Zn$ for the
overall ionic charge density.   This particular linear combination
was chosen, because (i) the charge density is the physical quantity
of interest here, and (ii) the electric
potential is decoupled from the ``charge-neutral'' density.

It is straightforward to solve the
``hard-core'' linear combination, Eq.~\eqref{el2a}. Imposing that
$\rho_+({\vec r})/\bar{\rho}_++\rho_-({\vec r})/\bar{\rho}_-\equiv
0$ within the hard-core of any of the colloids (i.e.\ wherever
$W({\vec r})=w_0$) yields a value for the hard-core parameter,
\begin{equation}
\beta w_0=\frac{1}{1-\eta}, \label{w0}
\end{equation}
whereas outside any of the colloidal hard core positions we have
\begin{align}\label{el2a2}
\frac{\rho_+(\vec r)}{\bar\rho_+}+\frac{\rho_-(\vec
r)}{\bar\rho_-} = \frac{2}{1-\eta}.
\end{align}

The solution of the ``charge'' linear combination,
Eq.~\eqref{el2b}, is most straightforwardly found by Fourier
transformation.  For an arbitrary function $f({\vec r})$ we define
and denote the Fourier transform as $f_{\vec k}=\int \d{\vec
r}f({\vec r})\exp(i{\vec k}\cdot{\vec r})$. One easily checks from
equation~\eqref{el2b} that
\begin{align}\begin{split}
\label{eq:rho_fourier}
\fourier\rho =&
  (2\pi)^3\left\{\bar\rho+(\bar\rho_++\bar\rho_-)\bar\phi
  +\bar{\rho}\eta\beta w_0\right\} \delta(\vec k) \\
&- (\bar\rho_++\bar\rho_-)\fourier\phi -\bar{\rho}W_{\vec k},
\end{split}\end{align}
where we have from Eq.~\eqref{w} that
\begin{eqnarray}\label{eq:fourierW}
\fourier W=\frac{4\pi a
w_0}{k^2}\left(\frac{\sin(ka)}{ka}-\cos(ka)\right)
\sum_{j=1}^N e^{i{\vec k}\cdot {\vec R}_j},
\end{eqnarray}
and from Eqs.~\eqref{eq:phi} and~\eqref{v} that
\begin{align}\begin{split}
\label{eq:phi_fourier} \fourier\phi =&
  4\pi\lB\frac{\fourier\rho}{k^2}
  -\frac{4\pi a}{k^2}
  \sum_{j=1}^N \exp(i\vec k\cdot\vec R_j)\\
&\times
    \left\{\left(\beta v_0 + Z\frac{\lB}{a}\right) \cos ka
    - \beta v_0 \frac{\sin ka}{ka}\right\}.
\end{split}\end{align}
Equations~\eqref{eq:rho_fourier} and~\eqref{eq:phi_fourier} are
two linear equations in the unknowns~$\phi_{\vec k}$
and~$\rho_{\vec k}$, which can be solved straightforwardly. Fixing the
remaining hard-core parameter $v_0$ to
\begin{align}\label{eq:v0}
\beta v_0 = -Z\frac{\bar\kappa\lB}{1+\bar\kappa a} + \beta
w_0\frac{\bar\rho_+-\bar\rho_-}{\bar\rho_++\bar\rho_-},
\end{align}
we find that the charge density is given by
\begin{align}\label{eq:rho_sol}
\begin{split}
\fourier\rho =
 & (2\pi)^3 \left\{\frac{\bar\rho}{1-\eta}+(\bar\rho_++\bar\rho_-)\bar\phi\right\}
   \frac{k^2}{k^2+\bar\kappa^2} \delta(\vec k) \\
 & +\frac{Z}{1+\bar\kappa a}
    \frac{\cos ka+\frac{\bar\kappa}{k}\sin ka}
         {1+k^2/\bar\kappa^2}
    \sum_j e^{i\vec k\cdot\vec R_j},
\end{split}
\end{align}
where the effective Debye screening parameter is defined as
\begin{align}\begin{split}\label{kappabar}
\bar\kappa\equiv&\sqrt{4\pi\lB(\bar\rho_++\bar\rho_-)}\\
=&\sqrt{4\pi\lambda_B}\sqrt[4]{(Zn)^2+(2\cs\exp[-\eta/(1-\eta)])^2}.
\end{split}\end{align}
Here we used Eqs.~\eqref{donnan}, \eqref{eq:salt}, and~\eqref{w0}
in rewriting the first into the second line. Note that the factor
$\exp[-\eta/(1-\eta)]$ that appears in Eq.~\eqref{kappabar} can be
accurately represented by $(1-\eta)$, with a relative deviation
less than $0.01$ for $\eta<0.1$ and less than~$0.1$ for $\eta<0.35$.

\begin{figure*}
\null\hfil
\subfigure[]{\epsfig{file=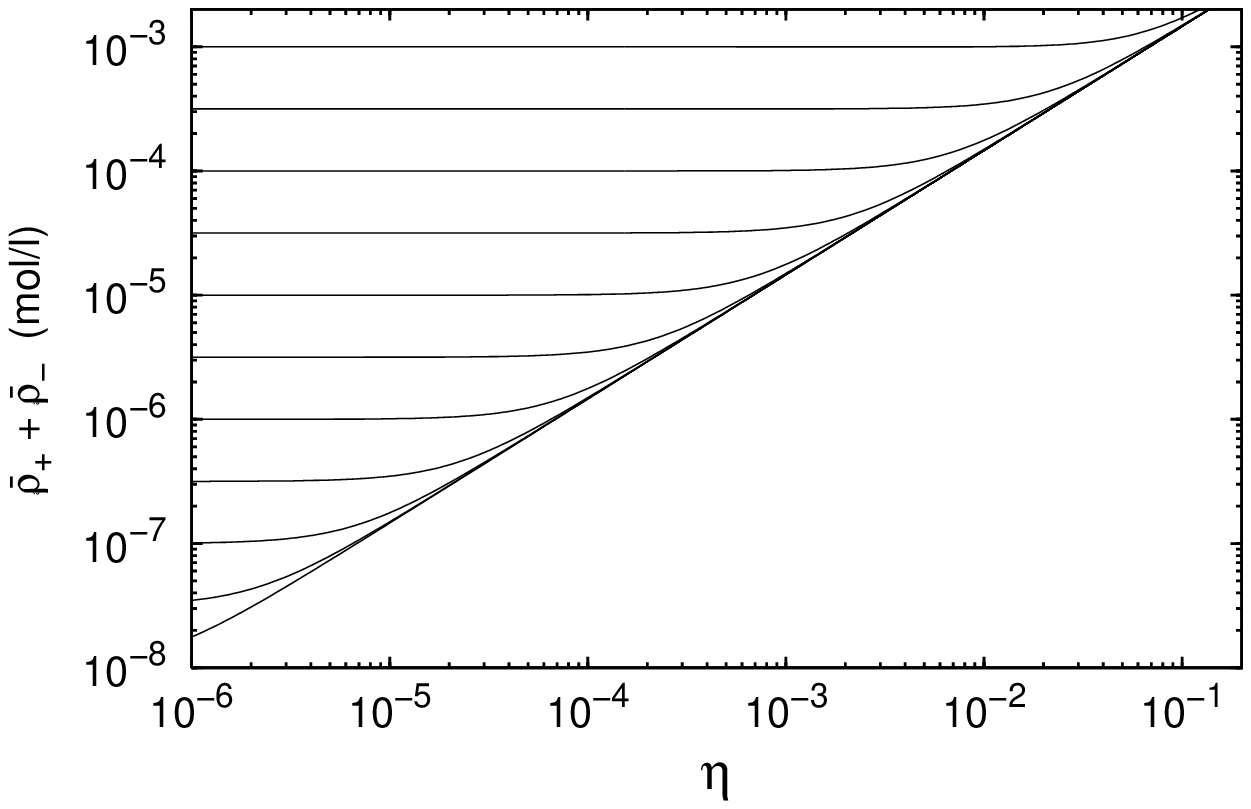,width=.47\hsize}}
\hfil
\subfigure[]{\epsfig{file=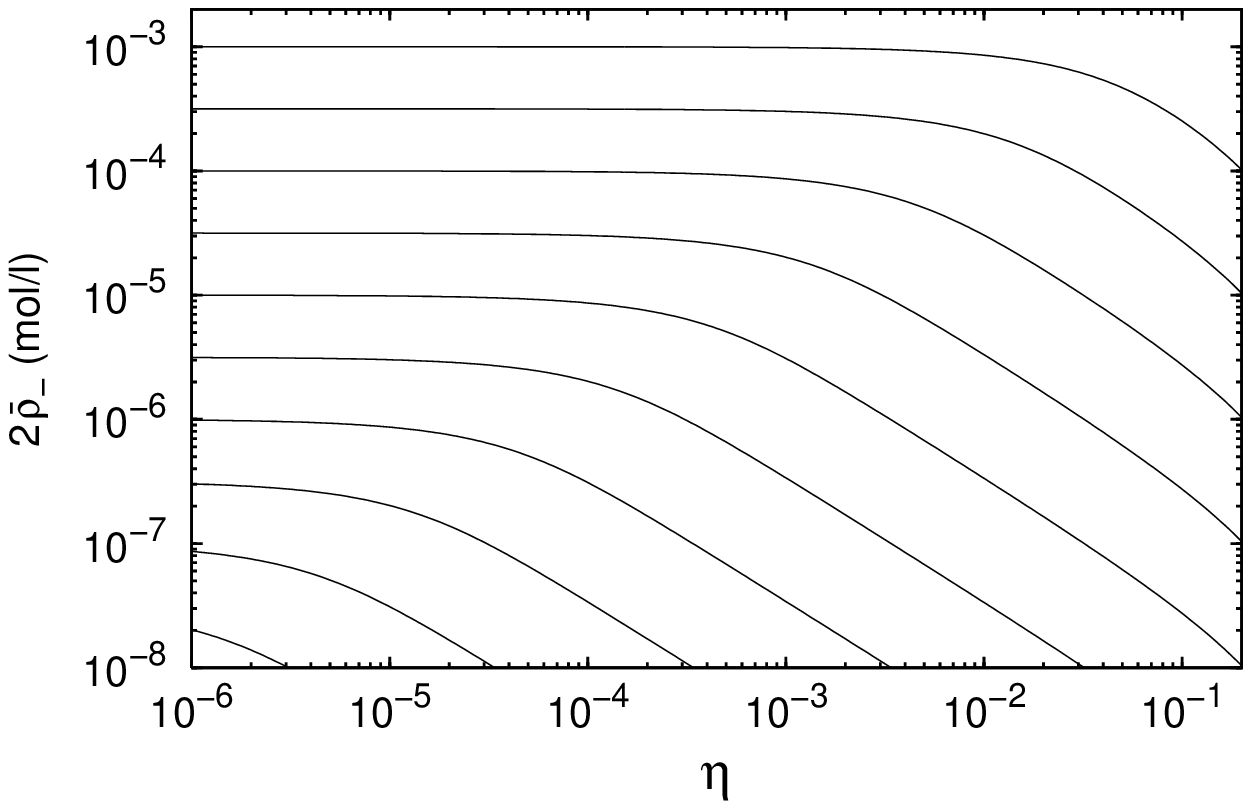,width=.47\hsize}}
\null\break
\caption{\label{fig:salt}
Total ion concentration~$\bar\rho_++\bar\rho_-$ (a) and
concentration of added salt~$\bar\rho_-$ (b) as a function of the
colloid packing fraction for different reservoir concentrations,
using the expressions of Eq.~\eqref{eq:salt} and~\eqref{w0}. The
colloidal charge and radius are $Z=50$ and $a=21.9\nm$,
respectively, and the solvent is ethanol at room temperature such
that $\lB=2.37\nm$. This matches the parameters from the
experiments by Ra{\c s}a et al~\cite{mircea}.}
\end{figure*}

The first term in expression~\eqref{eq:rho_sol} is of the
form $\propto k^2\delta({\bf k})$ and does {\em not} contribute to
the charge density~\eqref{eq:charge1}. However, we will see below
that this term does in fact contribute to the grand potential, as
this involves the Coulomb energy
$\propto\int \d{\vec k}\rho_{\vec k}/k^2$.

The real space
representation of the charge density is a multi-centered sum
$\rho(\vec r)=\sum_i\rho_1(|\vec r-\vec R_i|)$, where the
one-particle density profiles (the ``orbitals'') have the usual
DLVO form\cite{dl1941,vo1948}:
\begin{align}\label{eq:charge1}
\rho_1(r) = \begin{cases} 0 & r<a; \\ \displaystyle
 \frac{Z\bar\kappa^2}{4\pi}\frac{\exp(\bar\kappa a)}{1+\bar\kappa a} \frac{\exp(-\bar\kappa r)}{r}
 & r>a.
 \end{cases}
\end{align}
We note that the vanishing of $\rho_1(r)$  inside the colloidal
hard core is a direct consequence of our particular choice for
$v_0$ given by Eq.~\eqref{eq:v0}; other choices for $v_0$ would
have yielded a finite ion charge density inside the hard core.
Note also that the multi-centered charge density $\rho({\vec r})$
is {\em not} vanishing within the hard cores, since the
exponential tail of the orbital centered around colloid~$i$
penetrates the hard core of all the other colloids $j\neq i$.

By inserting Eq.~\eqref{w0} into \eqref{eq:salt}, explicit
expressions for the average concentrations~$\bar\rho_\pm$ of ions
in the suspension are obtained as a function of the colloid
density $n$, colloid charge $Z$, and the reservoir concentration
$2{\cs}$ --- this was already used to obtain Eq.~\eqref{kappabar}.
These expressions reduce, in the limit of point-like colloids (for
which $\eta=0$) to the standard
expressions for the Donnan effect~\cite{donnan1924,overbeek1956}.

This effect is illustrated in Fig.~\ref{fig:salt}, where we plot
the total ion concentration $\bar\rho_++\bar\rho_-$ in~(a), and
the concentration of added salt
$2\bar\rho_-=\bar\rho_++\bar\rho_--Zn$ in~(b), on the basis of our
expressions for $\bar{\rho}_{\pm}$. The parameters are close to
those of the experiments by
\citeauthor{mircea}~\cite{mircea,mircea_nature}: $Z=50$,
$\lB=2.37\nm$, and $a=21.9\mbox{nm}$. The reservoir salt
concentration equals the $\eta=0$ limit of each of the curves, and
the crossover from the low-$\eta$ plateau to the high-$\eta$
linear part corresponds to the crossover from added-salt dominance
to counterion dominance. Note the expulsion of added salt back
into the reservoir at high $\eta$ in~(b). An important aspect of
these intermediate results is that the screening parameter
$\bar{\kappa}$ {\em increases} with~$n$ essentially
$\propto\sqrt{Zn}$ in the counterion-dominated regime (which may
occur at packing fractions as low as $\eta\simeq 10^{-4}$ if
${\cs}\simeq 3\mM$).

As we have now solved the Euler--Lagrange equations \eqref{el1}
for the two linear combinations $\rho_+({\vec
r})/\bar{\rho}_++\rho_-({\vec r})/\bar{\rho}_-$ and $\rho_+({\vec
r})-\rho_-({\vec r})$, it is straightforward to disentangle the
equilibrium profiles and obtain the profiles~$\rho_\pm({\vec r})$
of the two ionic species separately.

\medskip

It is important to realize, however, that these results depend on
the particular choice that we have made for the hard-core
potentials in Eqs.~\eqref{v} and~\eqref{w}.  Different choices
for these hard-core potentials lead to other, non-equivalent
minima of the grand-potential. For instance, instead of
$U_\pm(\vec r)=\pm V(\vec r)+W(\vec r)$, we could have considered
the choice $U_\pm(\vec r)=\pm V(\vec r)+2\bar\rho_\mp W(\vec
r)/(\bar\rho_++\bar\rho_-)$, which, with $\beta
v_0=-Z\bar\kappa\lB/(1+\bar\kappa a)$ and $\beta w_0=1/(1-\eta)$
would lead to a vanishing $\rho_1(r)$ and $\rho_+({\vec
r})/\bar{\rho}_++\rho_-({\vec r})/\bar{\rho}_-$ inside the hard
cores.   This choice was actually made in
Refs.~\cite{PRL-79-3082,PRE-59-2010}, and leads to similar, but
not identical results --- see Appendix~\ref{appB}.

\subsection{The minimum of the functional}
\noindent 
In Appendix~\ref{appendix}, we derive the equilibrium grand
potential~$\Omega$ by insertion of our solution of the
Euler--Lagrange equations into the functional.  The effective
interaction Hamiltonian $H={\cal H}_{\rm cc}+\Omega$ then takes
the form
\begin{equation}\label{H2}
\begin{split}
H(\{{\vec R}\},N,V,{\cs}) =& 
  \Phi(V,n,{\cs})
+ \sum_{i<j}^N V(R_{ij};n,{\cs}).
%&+\PhiD(N,V,{\cs}) + \Phi_0(N,V,{\cs}),
\end{split}
\end{equation}
The first term $\Phi$, is independent of the colloidal
coordinates ${\vec R}_i$, and is called the ``volume term'' as it
is a density-dependent, extensive thermodynamic quantity that scales
with the volume of the system. The second term of Eq.~\eqref{H2} is
a pairwise sum that does depend on the colloidal coordinates (and
on the density $n$). For later convenience we decompose the
volume term as $\Phi=\PhiD+\Phi_0$, with the so-called ``Donnan''
term defined by
\begin{align}\label{eq:donnanterm}
\frac{\beta\PhiD}{V} = \sum_\pm
\bar\rho_\pm\left(\ln\frac{\bar\rho_\pm}{{\cs}}-1\right),
\end{align}
and the other term by
\begin{align}\label{eq:volterm1}
\begin{split}
\frac{\beta\Phi_0}{V} &=
-\frac{1}{2}\frac{(Zn)^2}{\bar\rho_++\bar\rho_-}
+\frac{\eta}{1-\eta}\frac{2\bar\rho_+\bar\rho_-}{\bar\rho_+ +
\bar\rho_-} -\frac{n}{2}\frac{Z^2\bar\kappa\lB}{1+\bar\kappa a}.
\end{split}
\end{align}
In Section~\ref{sec:osmotic} below, we will see that $\PhiD$,
which takes the form of ideal-gas contributions, accounts for the
Donnan equation of state (except for the colloidal ideal gas
contribution); hence the nomenclature. The term $\Phi_0$ appears
as an electrostatic (and hard-core) free energy contribution. This
separation is slightly misleading, however, since the two terms
both depend on $n$ and $Z$ through the expressions~\eqref{eq:salt}
and~\eqref{kappabar}, which stem from the Donnan potential~\eqref{donnan}
and hence from the balance between electrostatics
and entropy.

The effective pair potential between the colloids,
$V(R_{ij})$,  that appears in the second term of Eq.~\eqref{H2}, is
given by
\begin{equation}\label{V2}
\beta V(r)=\begin{cases}
  \infty & r<2a; \\ \displaystyle
  %\left( Z\frac{\exp(\bar\kappa a)}{1+\bar\kappa a}\right)^2
  (1+A)
  \Zdlvo^2
  \lB
  \frac{\exp(-\bar\kappa r)}{r} & r>2a,
\end{cases}
\end{equation}
with the DLVO-charge given by~$\Zdlvo=Z\exp(\bar\kappa
a)/(1+\bar\kappa a)$, and with the parameter $A$ defined by
\begin{align}\label{A}
A = 4\pi\beta w_0 \frac{\np}{\bar\kappa^3}
\left((1+\bar\kappa a)^2e^{-2\bar\kappa a}+(\bar\kappa a)^2 -1\right).
\end{align}
The effective pair interaction~$V(r)$ is very similar to the
traditional DLVO potential~$V_2(r)$ of Eq.~\eqref{V2gen}, but with
two important differences. The first difference involves the
screening parameter $\bar{\kappa}$ in $V(r)$, which is to be
contrasted with the reservoir screening parameter $\kappa$ in
$V_2(r)$. The second difference is that the amplitude of $V(r)$ is
enhanced compared to $V_2(r)$ by a factor $(1+A)$.  This can be
traced back to our particular choice of linear combinations of
density profiles that we used to solve the Euler--Lagrange
equations.

\begin{figure}
\epsfig{file=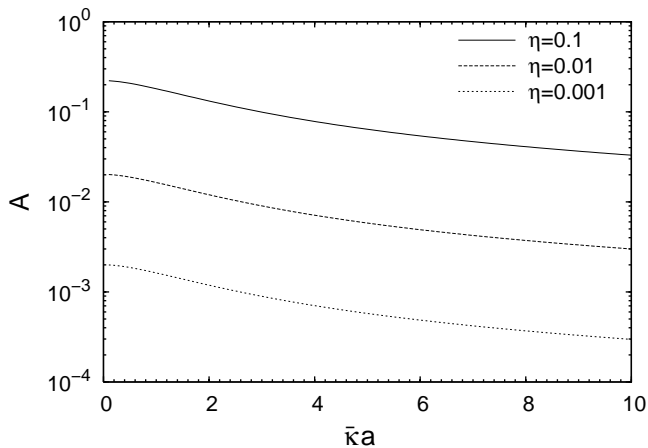,width=.99\hsize}%
\caption{%
The factor~$A$ of Eq.~\eqref{A} as a function of $\bar\kappa a$ and
$\eta$.  For all parameters, $A\ll1$, so it can safely be neglected in
Eq.~\eqref{V2}.%
}
\label{fig:ratio}
\end{figure}

In Figure~\ref{fig:ratio} we plot $A$ as a function of the
screening parameter $\bar\kappa a$ for several values of the
packing fraction $\eta$.  The plot shows that $A\ll 1$ for
essentially all packing fractions of interest here. Moreover, one
can also show that $A\equiv 0$ if the hard-potentials are defined
as $U(\vec r)=V(\vec r)+2\bar\rho_\mp W(\vec
r)/(\bar\rho_++\bar\rho_-)$ instead of the definition used here.
This latter choice does not affect any of the other volume terms,
but does involve another choice for $v_0$ and $w_0$, and does
change the expression of $\bar\phi$ (see Appendix~\ref{appB}).
For these two reasons we set $A\equiv 0$ in the remainder of the
paper.

The so-called volume terms $\PhiD$ and $\Phi_0$  are very similar
to their canonical counterparts that were derived in
Ref.~\cite{PRE-59-2010}. The main difference is that the present
volume term includes the term $-\int \d{\vec
r}\big(\mu_+\rho_+({\vec r})+\mu_-\rho_-({\vec r})\big)$ due to
the grand-canonical character of our calculations. This leads to
another difference, since one should now view the Hamiltonian~$H$
as a function of~$n$ and the reservoir salt concentration~${\cs}$,
i.e.\ one should take the dependence of~$\Phi$ on~$\bar\rho_\pm$
and~$\bar{\kappa}$ as a dependence on~${\cs}$ and~$n$ through
Eqs.~\eqref{kappabar} and~\eqref{eq:salt}. It is the nontrivial
(and nonlinear) dependence of $\beta\PhiD/V$ and $\beta\Phi_0/V$ on
the colloid density~$n$, at fixed~${\cs}$, that is responsible for
interesting thermodynamic effects, as we will see below.

%It may, at first sight, be uncomforting that the effective
%pair-interaction $V(r)$ that we obtained from our analysis depends
%on the density $n$ of colloids. Our hand-waving interpretation is
%that this $n$-dependence represents the effect of three-body
%potentials and higher-order terms within linearized
%Poisson--Boltzmann theory. \bla{is this true?  doesn't it simply
%represent the screening by the counterions?}

\section{Thermodynamics}
\subsection{Free energy}
\noindent
As we have now found the functional form~\eqref{H2} for the
effective Hamiltonian of the colloids, we are ready to calculate
the corresponding free energy $F(N,V,T,{\cs})$ defined just below
Eq.~\eqref{H}. From this the other thermodynamic quantities follow.
Since the volume terms in~\eqref{H2} are independent of the
coordinates of the colloids, we can factor out their Boltzmann weights
and write
\begin{equation}
\label{FF}
\exp(-\beta F)=
  \exp[-\beta\Phi]
  \tr_{\rm c}\exp\left[-\beta \sum_{i<j}^NV(R_{ij})\right].
\end{equation}
This can be rewritten as
\begin{align}
\label{Ftot} F = \PhiD + \Phi_0 +F_{\text{id}} + F_{\text{exc}},
\end{align}
with $\PhiD$ and $\Phi_0$ defined in Eq.~\eqref{eq:volterm1},
with the colloidal ideal-gas free energy
\begin{equation}
\label{Fid} F_{\text{id}}=N{\kB}T\big(\ln(n{\cal V})-1\big),
\end{equation}
and where $F_{\text{exc}}$ is the non-ideal (excess) free energy
due to the colloid-colloid pair interactions \eqref{V2}. Here we
calculate $F_{\text{exc}}$ variationally, using the
Gibbs--Bogoliubov inequality~\cite{HMcD,Isihara,Gibbs,Bogoliubov}.
This inequality states that the excess free energy
$F_{\text{exc}}^{\text{(ref)}}$ of a so-called reference system of
volume~$V$ that contains~$N$ particles with {\em any} pair
interaction $V^{\text{(ref)}}(R_{ij})$, satisfies
\begin{align}\label{eq:gibbs-bogoliubov}
F_{\text{exc}} \leq F_{\text{exc}}^{\text{(ref)}}
+\left\langle
 \sum_{i<j}\left(V(R_{ij})-V^{\text{(ref)}}(R_{ij})\right)
\right\rangle_{\text{ref}},
\end{align}
where $\langle\dots\rangle_{\rm ref}$ denotes a thermodynamic
average that is to be evaluated in the reference system. The key
idea is to use a reference pair potential with a free parameter
with respect to which the right hand side of
Eq.~\eqref{eq:gibbs-bogoliubov} can be minimized; the minimum is
then the optimal estimate for the free energy $F_{\text{exc}}$ of
interest. We use two different reference system to calculate the
free energy of fluid and crystal phases, respectively.

For the fluid phase we use a hard-sphere reference system, with
the hard-sphere diameter $d$ as variational parameter. Introducing
the effective hard-sphere packing fraction $\xi=(\pi/6)nd^3$, we
can write
\begin{equation}
\label{fexcfluid}
\begin{split}
\frac{F_{\text{exc}}}{N{\kB}T}=\min_d \biggl\{
 &  \frac{4\xi-3\xi^2}{(1-\xi)^2} \\
 &+ \frac{n}{2}4\pi\int_d^{\infty}\d{r} r^2 g_d(r;\xi) \beta V(r)
 \biggr\},
\end{split}
\end{equation}
where the first term is Carnahan--Starling expression for the
hard-sphere free energy~\cite{HMcD,JCP-51-635}, and where
$g_d(r;\xi)$ is the radial distribution function of a fluid of
hard spheres with diameter $d$ and packing fraction $\xi$. We
approximate $g_d(r;\xi)$ by the Verlet--Weis corrected
Percus--Yevick expressions \cite{PRA-5-939,HMcD}. This allows for
an analytic evaluation of the integral in Eq.~\eqref{fexcfluid},
since the Yukawa form of $V(r)$ turns this integral into a Laplace
transform of $r g_d(r;\xi)$, for which accurate expressions have been
derived on the basis of Pad\'e fits in
Refs.~\cite{PRA-43-5418,PRE-53-4820}. The minimization with
respect to $d$ is then numerically performed straightforwardly.
Note that such a minimum indeed exists, as the excess free energy
of the hard-sphere reference system becomes infinitely large in
the limit of large particle sizes.  Because the particles in our
actual system have also a hard core of radius $a$, we impose that
$d\geq2a$.

As a reference system for the solid phase, we use $N$~classical
Einstein oscillators~\cite{ziman,ashcroft} on an FCC
lattice~\cite{JCP-86-5127}.  The Einstein
frequency~$\omega_{\text{E}}$ plays the role of the variational
parameter used to minimize the right hand side of
Eq.~\eqref{eq:gibbs-bogoliubov}.  For the thermodynamic average of
the Yukawa interactions in this system, we use the expressions
found by \citeauthor{JCP-86-5127}~\cite{JCP-86-5127}.

So far we have only considered FCC configurations for the solid
phase, but there is no principal problem to generalize this to
other structures such as BCC \cite{PRE-59-2010}. Consequently, we
only consider gas--liquid, fluid--FCC and FCC--FCC phase
equilibriums in this paper.

\subsection{Osmotic pressure}\label{sec:osmotic}
The osmotic pressure $\Pi=P-2{\cs}kT$ of the suspension under
consideration follows from $P=-\partial F/\partial V$ at fixed~$N$
and~${\cs}$. We can therefore use our expression for $F$ given in
Eq.~\eqref{Ftot} to obtain $\Pi(n,{\cs})$ explicitly as
\begin{equation}
\label{pres} \Pi=\PiD+\Pi_0+\Pi_{\text{id}}+\Pi_{\text{exc}}
\end{equation}
with ${\PiD}=-2{\cs}\kB T-(\partial \PhiD/\partial V)$, the Van
't Hoff (ideal-gas) contribution $\Pi_{\text{id}}=-(\partial
F_{\text{id}}/\partial V)=n{\kB}T$, the excess pressure
$\Pi_{\text{exc}}=-(\partial F_{\text{exc}}/\partial V)$, and the
remaining term $\Pi_0=-(\partial \Phi_0/\partial V)$. Explicit
general expressions for ${\PiD}$, $\Pi_0$, and $\Pi_{\text{exc}}$
can of course be given on the basis of Eqs.~\eqref{eq:volterm1},
\eqref{eq:donnanterm}, and e.g.~\eqref{fexcfluid}, respectively,
but it turns out to be instructive to focus on these expressions
in the limit of point-colloids with radius $a=0$ (such that
$\eta=0$): this reduces the algebra and allows for an interesting
illustration of cancellations of some of the electrostatic
contributions to the osmotic pressure~$\Pi$. We stress, however,
that we used the full expressions in our numerical calculations
presented below.

In the point-colloid limit we have
\begin{align}\begin{split}\label{pd}
\beta {\PiD} =& -2\cs + \bar{\rho}_+ + \bar{\rho}_-\\
=&-2\cs+\sqrt{(Zn)^2+(2\cs)^2}\\[6pt]
=&\begin{cases}
    \displaystyle\frac{(Zn)^2}{4\cs} + {\cal O}(n^4) & Zn\ll 2\cs\\
    \displaystyle Zn-2\cs+{\cal O}(\cs^2) & Zn\gg 2\cs,
  \end{cases}
\end{split}\end{align}
and a little tedious but straightforward algebra yields
\begin{eqnarray}\label{p0}
\beta\Pi_0=\begin{cases}
  \displaystyle -\frac{(Zn)^2}{4{\cs}}+ {\cal O}(n^3) & Zn\ll 2\cs\\[6pt]
  \displaystyle -b n^{3/2} + {\cal O}(\cs^2) & Zn\gg 2\cs,
\end{cases}
\end{eqnarray}
with a coefficient $b=\sqrt{Z\pi\lambda_B}Z^2\lambda_B/2$.  We focus
first on the low-density/high-salt regime $Zn\ll 2\cs$, and then on
the opposite regime.

The expressions~\eqref{pd} and~\eqref{p0} show a cancellation of
the dominant term in the regime $Zn\ll 2{\cs}$, such that in this
regime $\Pi\simeq \Piid+\Pi_{\text{exc}}$, i.e. the pressure is
actually the pressure of the effective one-component system
described by the pairwise screened-Coulomb Hamiltonian.
Interestingly, however, one can also write the virial expansion
$\beta F_{\text{exc}}/V=B_2(\bar{\kappa})n^2+{\cal O}(n^3)$  in this
regime, where the second virial coefficient~\cite{HMcD} is
\begin{equation}
\label{vir} B_2(\bar{\kappa})=\frac{1}{2}\int\d{\vec r}
\big(1-\exp[-\beta V(r)]\big),
\end{equation}
with the colloidal pair potential~$V(r)$ defined in Eq.~\eqref{V2}.
In the limit of weak interactions, the exponent in Eq.~\eqref{vir} can
be linearized with the result that $B_2=Z^2/4\cs$ for
point-colloids. This means that $\beta\Pi_{\text{exc}}\simeq
(Zn)^2/4\cs\simeq\beta{\PiD}$, and hence that the pressure can also
be approximated by the Donnan expression $\Pi\simeq \Piid+{\PiD}$.
In other words, on the basis of this simple analysis
one expects ``reliable'' results for the pressure (and hence for the
thermodynamics) in the regime $Zn\ll 2\cs$  by taking either the full
four-term expression~\eqref{pres} for~$\Pi$, or the two two-term
expressions $\Piid+\PiD$ and $\Piid+\Pi_{\text{exc}}$, but {\em not}
any other combination.  This
will be confirmed by our numerical results below.

\medskip

The situation is bit more complicated in the opposite low-salt
regime $2{\cs}\ll Zn$, since then (i) no cancellations take place and
(ii) the virial expansion for $F_{\text{exc}}$ breaks down because
of the long-range character of the unscreened-Coulomb interactions.

%One can show, e.g.\ by setting $g_d(r)\simeq \exp(-\beta V(r))\simeq
%1-\beta V(r)$ with $\bar{\kappa}=\sqrt{4\pi\lambda_BZn}$ and $d=0$
%in Eq.~\eqref{fexcfluid}, that the lowest order contribution to the
%excess pressure takes the Debye--H\"uckel-like form
%$\beta\Pi_{\text{exc}}=- b'n^{3/2}$ with
%$b'=\sqrt{\pi}Z^2(Z\lambda_B)^{3/2}/4$. As a consequence we find the
%asymptotic low-salt result

As a simple approximation for highly charged particles
(specifically, particles for which $Z^2\lB/a\gg1$), the pair
correlation function~$g_d(r)$ can be set to $g_d(r)=1$ for
$r\gtrsim \frac{1}{2}n^{-1/3}$ and to $0$ otherwise.  One can then show
that the lowest order contribution to the excess pressure takes the
form $\beta\Pi_{\rm exc}=-b' n^{4/3}$ with $b'=\pi Z^2\lB/12$. As
a consequence we find the asymptotic low-salt result
\begin{equation}\label{plowsalt}
\beta\Pi=(1+Z)n - b'n^{4/3} - bn^{3/2},
\end{equation}
which contains Donnan, colloidal pair, and Debye-H\"uckel-like
contributions. The prefactors of the fractional powers would
change if a proper G\"untelberg charging process would have been
performed \cite{JCP-113-9722}, but the present analysis is good
enough to capture the spinodal instability that is now well-known
to be realistic for primitive model systems at sufficiently strong
coupling (low enough temperature)
\cite{antti-preprint,PRL-71-3826,JCP-107-1565,JCP-110-1581,
JCP-111-9509,levin_fisher,PRL-88-045701}. On this basis one could
expect that the present theory predicts phase-separation in
low-salt colloidal suspensions. Within the full theory for $F$ we
indeed find this phenomenon in the next section.

%Note that this phase transition
%only comes about when all four contributions to $F$ in
%Eq.~\eqref{Ftot} are taken into account; unlike the high-salt regime
%there is no cancellation effect in the low-salt regime.  \bla{is
%this true?  I think $\Phi_0$ by itself is enough to drive a
%phase-separation.}

\medskip

\begin{figure}[t]
\epsfig{file=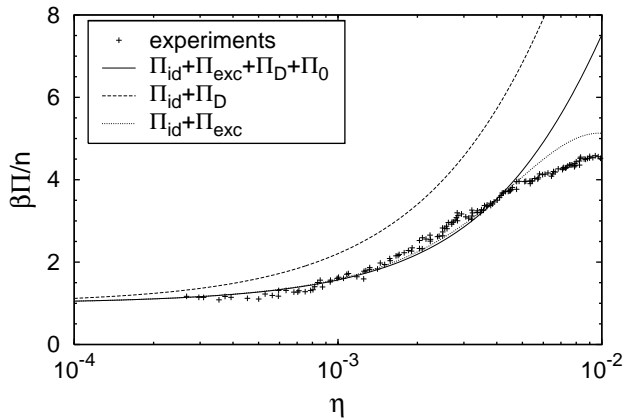,angle=-90,width=.99\hsize} \caption{Equation of
state compared to the experiments of Ra{\c s}a et al\cite{mircea},
described by the parameters $Z=32$ for the colloidal charge,
$a/\lB=96.7$ for the radius-to-Bjerrum length ratio, and
$2\cs=16~\mM$ for the reservoir salt concentration. Shown are the
experimental data (crosses), a one-component DLVO system (dotted),
the Donnan theory (dashed), and our full linear theory (solid
line). Two of the three theoretical curves describe the
experimental curves accurately for $\eta\lesssim0.003$, the Donnan
theory is less accurate although still qualitatively reasonable in
this regime.} \label{fig:mircea}
\end{figure}

We now illustrate our results for the osmotic pressure by
numerically comparing the theoretically predicted values to
experimental measurements in Figure~\ref{fig:mircea}. The
experimental system is an ethanol suspension of colloidal
silica-spheres, for which $\Pi(n)$ was determined by integration
of the measured density profile in sedimentation equilibrium
\cite{mircea_nature,mircea}.  The system parameters are $Z=32$,
$2{\cs}=16~\mM$, $\lB=2.38~\nm$\ and $a=21.9~\nm$. Since
$Z\lB/a\approx3$ we do not expect too much charge
renormalization, and as $Zn/2{\cs}\approx0.7$ at the highest
density considered here ($\eta=0.01$), this experiment is expected
to be in the high-salt regime where not only the full expression
\eqref{pres} for $\Pi$ but also both the one-component expressions
$\Pi\simeq \Piid+\Pi_{\text{exc}}$ and the Donnan expression
$\Pi\simeq \Piid+{\PiD}$ are expected to ``work'' with reasonable
accuracy. 

This is to some extent confirmed by
Fig.~\ref{fig:mircea}, where the measured osmotic pressure is in
quantitative agreement with two of the three theoretical versions
at low packing fractions $\eta\lesssim0.003$ or so; the Donnan
pressure is less accurate. At higher densities the different
theoretical curves deviate from each other (and from the
experiment), with the one-component result
$\Piid+\Pi_{\text{exc}}$ being closest to the actual experiment. A
word of caution is appropriate here, however, since recent work by
Biesheuvel indicates that charge regularization is relevant in the
present system, i.e.~the bare colloidal charge $Z$ is {\em not} a
constant but decreases with density, where significant deviations
of the low-density charge is predicted for $\eta\gtrsim 0.002$
\cite{JPCM-16-L499}. This is rather precisely the regime where the
theories begin to deviate from the experiment. This issue will
also be addressed in more detail in future work.

From the fact that the one-component osmotic pressure
$\Pi=\Piid+\Pi_{\text{exc}}$ describes the experimental data
rather accurately, one may conclude that the experimentally found
``inflated'' profiles of Ref.~\cite{mircea_nature} need not
necessarily be described by theories such as those of
Ref.~\cite{JPCM-15-S3569}, where a three-component mixture
(cations, anions, and colloids) in gravity gives rise to an
ion-entropy-induced self-consistent electric field that lifts the
colloids to higher altitudes than expected on the basis of their
mass. The equation of state suggests that an alternative
description could be given, based on hydrostatic equilibrium of an
effective one-component system of colloidal spheres with pairwise
screened-Coulomb repulsions only. The latter picture is {\em not}
in contradiction with the existence of the electric field, since
the density variation with height implies a variation of the
Donnan potential with height through Eq.~\eqref{donnan}. The two
pictures are, in this sense, merely two sides of the same coin, at
least on length scales beyond which the local density
approximation applies that underlies the one-component theory. On
smaller length scales the {\em source} of this electric field
involves deviations from local charge neutrality, which cannot be
explained by hydrostatic equilibrium and a bulk equation of state
alone.

\begin{figure}[t]
\epsfig{file=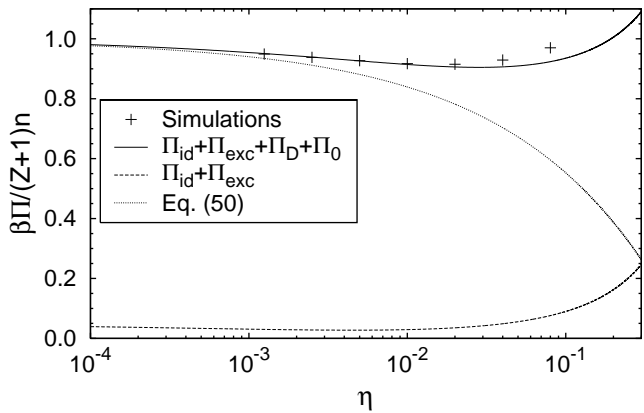,angle=0,width=.99\hsize} \caption{ Equation of
state (compressibility factor) compared to the computer simulation
of Ref.~\cite{JCP-113-4359}, where $Z=40$ and $a/\lB=22.5$. Shown
are the simulation data from~\cite{JCP-113-4359} (crosses), the
pressure $\Pi$ from our full linear theory (solid line), the
approximate low-salt expression \eqref{plowsalt} (dotted), and the
pairwise one-component result $\Piid+\Pi_{\text{exc}}$ (dashed).
The theoretical curves are based on the reservoir salt
concentration $\cs=10^{-15}~\mathrm{M}$, which is low enough to ensure an
essentially vanishing coion concentration in all state points
shown here.} \label{fig:linse}
\end{figure}
The other system for which we calculate the osmotic pressure is
one of the systems that Linse studied by Monte Carlo simulations
in Ref.~\cite{JCP-113-4359}. This system is free of added salt,
contains colloids with a charge $Z=40$ and a radius-to-Bjerrum
length parameter of $a/\lB=22.5$ for monovalent ions (in the
notation of Ref.~\cite{JCP-113-4359} the coupling parameter is
$\Gamma_{\rm II}=0.0445$). The simulated results are shown in
Figure~\ref{fig:linse}, together with several versions of the
present theory. It is clear that the major contribution to the
osmotic compressibility factor originates from the pressure
$\Piid+\PiD\approx(Z+1)n$, which exceeds the one-component
combination $\Piid+\Pi_{\text{exc}}$ by at least an order of
magnitude. The decrease of $\beta\Pi/n$ for $\eta\lesssim0.02$ is
due to the contribution $\Pi_0$. Our calculated pressure describes
the simulation data quite well, showing that volume terms may have
a pronounced effect on the thermodynamic properties of low-salt
suspensions, while the pairwise DLVO-picture without volume terms
breaks down qualitatively. We note, finally, that the limiting
expression~\eqref{plowsalt} for the pressure in the limit for
point-colloids can be seen to catch the low-density negative
curvature of $\beta\Pi/n$ with $n$ as predicted by the full theory
and the simulations, but {\em not} the increased stability at
higher $n$.

\section{Phase diagrams}
\noindent From the free energy per unit volume~$f(n)=F/V$ at fixed
${\cs}$, we calculate the chemical potential and the pressure, and
we impose the usual condition of thermodynamic
equilibrium~\eqref{grcancoex} to find a phase-equilibrium. We
already mentioned that this is numerically much less involved than
in the canonical calculations of e.g.\ Ref.~\cite{PRE-59-2010},
where the set~\eqref{cancoex} is to be solved. We merely
illustrate the feasibility of these calculations here by showing
two phase diagrams, for a particular~$Z$, $a$, and~$\lB$.
In forthcoming publications we will fully exploit the
relative simplicity of the grand-canonical formulation of the
theory by ``scanning'' the full parameter space $(Z,\lB/a)$,
including a generalization of the present theory to include charge
renormalization \cite{bas_metaphase}.

\begin{figure}
\epsfig{file=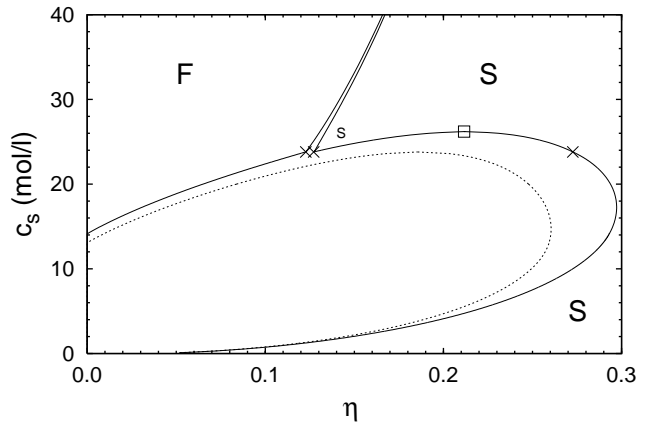,angle=-0,width=.99\hsize}
\caption{Phase diagram for a colloidal suspension as a function of
colloidal packing fraction~$\eta$ and reservoir salt concentration~${\cs}$.
The colloidal radius and charge are $a=326~\nm$ and
$Z=7300$, and the solvent is water at room temperature such that
the Bjerrum length is $\lB=0.72~\nm$. The solid lines denote
fluid--solid and solid--solid binodals, and the dotted line shows
the underlying metastable gas--liquid binodal.  The
fluid--solid--solid triple point is denoted by $\times$, and the
solid--solid critical point by $\square$.} \label{fig:phase1}
\end{figure}
The first set of parameters that we consider is $Z=7300$,
$a=326~\nm$, and $\lB=0.72~\nm$, which corresponds to the
experiments of Ref.~\cite{Nature-385-230}. The phase diagram that
follows from the present theory is displayed in
Figure~\ref{fig:phase1}, and shows phase coexistence with a
considerable or large density gap at ${\cs}\lesssim 20~\mM$, and
only a very small density gap at higher ${\cs}$. At a salt
concentrations of about $23.8~\mM$, a liquid--solid--solid triple
point occurs (denoted by $\times$ in Fig.~\ref{fig:phase1}), and
at $26.2~\mM$ a solid--solid critical point is located (denoted by
$\square$ in the figure).  Although somewhat difficult to see in
this picture, there is no lower critical point.

The phase diagram
of Fig.~\ref{fig:phase1} is pretty similar to the one calculated in
Ref.~\cite{PRE-59-2010} using the canonical version of the
theory~\cite{PRE-59-2010}, but with a few substantial differences.
The canonical theory, for instance, does not find any solid--solid
coexistence, nor does it find a triple point for these parameters.
Also the canonical theory predicts a lower critical point, while
the current grand-canonical version of the theory does not.
Despite this differences the main phenomenon is shared that at low
salinity ${\cs}\lesssim 20~\mM$ a density gap opens up.

The physical
mechanism for this demixing transitions into a dilute and dense
phase is identical to what was explained in
Refs.~\cite{PRE-59-2010,JPCM-11-10047,JPCM-12-A263}: the self
energy of the double layers as represented by the third term of
the volume term $\Phi_0$ in Eq.~\eqref{eq:volterm1} drives a spinodal
instability at low enough ${\cs}$, even though the pair interactions
are purely repulsive. The underlying physical mechanism is that
the cohesive energy that stabilizes the dense phase stems from the
compression of the double layers thickness $\bar{\kappa}^{-1}$
upon increasing the colloid density: this effect brings the charge
in the diffuse double layer closer to the oppositely charged
colloidal surface.  This mechanism is very similar to the one that
causes gas--liquid demixing in the restrictive primitive model
according to Debye--H\"uckel
theory~\cite{PRL-71-3826,JPCM-8-9103}.

A word of caution is appropriate here: given that
$Z\lB/a\simeq16$, one expects a substantial renormalization of the
charge within nonlinear Poisson--Boltzmann theory for this system,
and hence a reduction of the tendency to demix.  Whether or not
this mechanism for phase separation remains strong enough to yield
a big density gap in the phase diagram if charge renormalization
is taken into account, will be investigated in a future
publication~\cite{bas_multcel}.

\medskip

\begin{figure}[t]
\epsfig{file=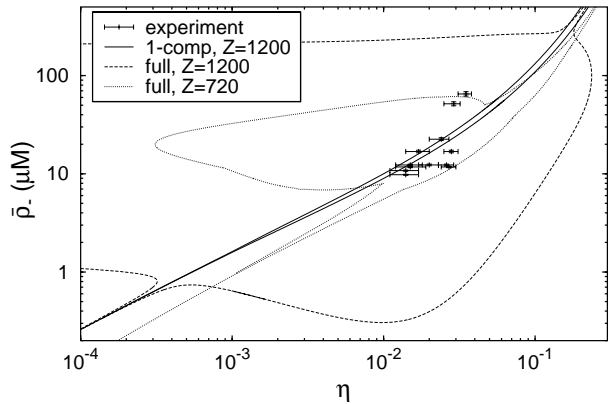,angle=0,width=.95\hsize}
\caption{%
Phase diagrams for the parameters of the experiment by
\citeauthor{JCI-128-533}~\cite{JCI-128-533}. The Bjerrum length
for this system is~$\lB=0.72~\nm$ and the partical radius
is~$a=66.7~\nm$.
The data points plotted here correspond to the samples
for which a fluid--solid coexistence was observed in Ref.~\cite{JCI-128-533}.
The solid line is a phase diagram
for a one-component DLVO system, the dashed line denotes the phase boundaries
for our full linear theory for a charge of $Z=1200$.  The dotted line gives
the phase boundary for the same full linear theory, but with a lower charge
$Z=720$.
}
\label{fig:phase2}
\end{figure}
The second phase diagram that we present here is for the
parameters of the experiments of
\citeauthor{JCI-128-533}~\cite{JCI-128-533}, where
$Z=1200$, $a=66.7~\nm$, and $\lB=0.72~\nm$. These parameters
were chosen because the experiments reveal a significant density
gap, by a factor of three,  between the coexisting fluid and solid
phases at salinity of the order of $10~\mM$. Such a large density
gap cannot be explained by the DLVO pair potential alone, and
hence we investigate here to what extent the volume terms may
account for this effect.

The phase diagram, shown in the $\eta$--$\bar\rho_-$
representation in  Fig.~\ref{fig:phase2}, shows the experimental
points and three fluid--solid binodals based on the present
theory.  As the results of Ref.~\cite{JCI-128-533} seem to be
independent of the concentration of added salt for salt
concentration lower than approximately $8~\mM$, we assumed an
extra background salt concentration of $8~\mM$ for the
experimental points.  Note that this representation of the phase
diagram, with the vertical axis representing the concentration of
added salt instead of the reservoir concentration, is such that
the tie-lines (which have been omitted for clarity) are no longer
horizontal as in the $\eta$--${\cs}$ representation, but instead
tilted to lower $\bar\rho_-$ at higher $\eta$ due to the Donnan
effect (see also Fig.~\ref{fig:salt}).

The first binodal in Fig.~\ref{fig:phase2} is the one based on the
ideal and excess part of~$F$ only, i.e.\ we assume that $\Phi_0$
and $\PhiD$ vanish (or more accurately: the volume terms are
assumed to be merely linear in~$N$ and~$V$ and do therefore not
affect the phase diagram). Although this binodal gives a fair
representation of the experimental points (probably this is how
$Z=1200$ was chosen), they do not capture the large density gap.
The second binodal is based on the full expression for~$F$,
including the volume terms, with $Z=1200$. We find an enormous
density gap that is much larger than experimentally observed, and
that extends to unreasonably high salt concentrations. The third
binodal is also based on our expression for $F$ with volume terms,
but now for a smaller charge $Z=720$. Interestingly, this choice
gives a density-gap at fluid--solid coexistence in the right salt
concentration regime, but the magnitude of the gap is yet much
bigger than experimentally observed. The reduction of the charge
from $Z=1200$ to $Z=720$ may give a rough idea of the effect of
charge renormalization, and shows that this nonlinear effect
reduces the tendency to demix considerably. Theories for charge
renormalization~\cite{JCP-80-5776,Lang-19-4027} show that the
renormalized charge is actually not a constant but depends on the
screening parameter and the density; the present value $Z=720$
corresponds to the dilute limit value at $\kappa a\approx0.8$,
i.e.\ at ${\cs}\approx10~\mM$, and is fixed here for simplicity. We
expect this to be a reasonable lower limit for the renormalized
charge in the region in which the phase-separation occurs. Also
this system will be investigated within a nonlinear version of
Poisson--Boltzmann theory in a future
publication~\cite{bas_multcel}.

\section{Conclusion}
\noindent We have reformulated and re-derived the volume term
theory for suspensions of charged colloids
\cite{PRE-59-2010,PRL-79-3082}. Our present derivation should be
more transparent than the original one, for instance because we
can now avoid the extra parameter $\lambda$ that regulates the
Coulomb potential from $1/r$ to $\exp(-\lambda r)/r$ with
$\lambda\rightarrow 0$ only at the end of the calculation.
Moreover, the presently derived expressions should be easier to
use in numerical calculations of thermodynamic properties and
phase diagrams, because the ions are treated grand-canonically
instead of canonically, thereby assuring equal chemical potential
of the ions from the outset. Moreover, a direct connection with
Donnan theory is now made, with explicit expressions for the
Donnan potential and the ion concentration in the system. In
future publications we will fully exploit the computational
advantages and extend the theory to include charge
renormalization.

We derived analytic expressions for the osmotic pressure in the
point-colloid limit for both the low-salinity and high-salinity
limits.  The low-salinity limit of the pressure was shown to
correspond to the Donnan expression, while in the limit of high
salt concentrations the traditional DLVO results are recovered.
The present full theory interpolates between these results, and
gives a good account of measured and simulated osmotic pressures
in both regimes.

We also calculated two phase diagrams.  The first one matches the
parameters of Ref.~\cite{PRE-59-2010}, and shows a similarly large
phase-instability at low salinity, although there are also a few
substantial differences.  The second phase diagram matches the
parameters of the experiments by \citeauthor{JCI-128-533} of
Ref.~\cite{JCI-128-533}, where an anomalously large density gap at
fluid--solid coexistence was reported.  Interestingly, the present
theory does predicts a density gap at fluid--solid coexistence, but
its magnitude is much larger than experimentally observed. We
stress, however, that these phase diagrams are calculated in a
regime where charge renormalization cannot be ignored. The relative
transparency of the present derivation allows to systematically
include this nonlinear effect into the theory, as will be shown in a
future publication~\cite{bas_multcel}.

The linear theory described in this paper already shows, however,
that volume terms can affect the osmotic pressure of low-salt
suspensions qualitatively, also in regimes where charge
renormalization and other nonlinear effects are {\em not} expected
to be important.

\section{Acknowledgement}
It is a pleasure to thank Mircea Ra{\c s}a and Albert Philipse for
collaborations and sharing their osmotic pressure data with us.
This work is part of the research program of the ``Stichting voor
Fundamenteel Onderzoek der Materie (FOM)'', which is financially
supported by the ``Nederlandse Organisatie voor Wetenschappelijk
Onderzoek (NWO)''.

%%%%%%%%%%%%%%%%%%%%%%%%%%%%%%%%%%%%%%%%%%%%%%%%%%%%%%%%%%%%%%%%%%
\appendix
\section{The grand potential}\label{appendix}
\noindent
In this Appendix, we derive the equilibrium grand
potential~$\Omega$.  We show that upon insertion of this grand
potential into Eq.~\eqref{H}, the effective Hamiltonian can be cast
into the form specified by Eqs.~\eqref{H2}--\eqref{V2}.

In the framework of Density Functional Theory, the equilibrium grand
potential is given by the minimum of the
functional~$\Omega[\rho_+,\rho_-]$ of Eq.~\eqref{eq:functional2}.
This minimum is found by inserting the Euler--Lagrange
equation~\eqref{el1} into the functional.  This
leads to the following expression for the grand potential:
\begin{align}\label{eq:grandpot_gen}
\begin{split}
\frac{\beta\Omega}{V} = &\sum_\pm
\bar\rho_\pm\left(\ln\left[\frac{\bar\rho_\pm}{{\cs}}\right]-1\right)
+\frac{Zn}{2}\bar\phi  \\
+&{\eta\beta w_0} \frac{\bar\rho_+ + \bar\rho_-}{2}
+\frac{1}{2V}\int\d{\vec r}\rho(\vec r)\beta V(\vec r) \\
+&\frac{1}{2V}\int\d{\vec r}\beta W(\vec r) \big(\rho_+({\vec r})
+\bar\rho_-({\vec r})\big).
\end{split}
\end{align}
The ``electrostatic'' integral can be evaluated as
\begin{align}\begin{split}
\label{eq:electroint}
\frac{1}{2V} & \int\d{\vec r}\rho(\vec r)\beta V(\vec r)
 = \frac{1}{2V}\frac{1}{(2\pi)^3}\int \d{\vec k} \beta V_{\vec k}\rho_{-{\vec k}}\\
=&
 - \frac{1}{2}\frac{1}{1-\eta}
   \frac{(\bar\rho_+-\bar\rho_-)^2}{\bar\rho_++\bar\rho_-}
 - \frac{Z\np}{2}\bar\phi
 - \frac{\np}{2}\frac{Z^2\bar\kappa\lB}{1+\bar\kappa a} \\
&+ \frac{1}{V} \sum_{i<j}\left\{
    (1+\frac{A}{2})\left(\frac{Ze^{\bar\kappa a}}{1+\bar\kappa a}\right)^2
    \frac{\lB e^{-\bar\kappa R_{ij}}}{R_{ij}} \right\} \\
&- \frac{1}{V} \sum_{i<j}\left\{
    - Z^2 \frac{\lB}{R_{ij}} \right\},
\end{split}\end{align}
where we inserted the Fourier transform $\rho_{\vec k}$ of $\rho(\vec r)$
from~\eqref{eq:rho_sol}, and the Fourier transform $V_{\vec k}$ of $V(\vec r)$,
which is given by
\begin{align}\begin{split}\label{eq:fourierV}
\beta\fourier V =
-\frac{4\pi}{k^3}\biggl\{
    &\left(\beta v_0+Z\frac{\lB}{a}\right) ka\cos ka \\
    &-\beta v_0 \sin ka
\biggr\}
\sum_j e^{i\vec k\cdot\vec R_j}.
\end{split}\end{align}
The factor~$A/2$ in the fourth term on the right hand side of
Eq.~\eqref{eq:electroint} is caused by the expulsion of microionic
charges from the colloid cores, and is given by Eq.~\eqref{A}.
Note that the first and second term of Eq.~\eqref{eq:electroint}
result from the $\propto k^2\delta({\vec k})$ term in
equation~\eqref{eq:rho_sol} that did not contribute to the charge
density.

\medskip

In a similar way, the ``hard-core'' part of the grand
potential~\eqref{eq:grandpot_gen} is evaluated as
\begin{align}\label{eq:hardcoreint}
\begin{split}
\frac{1}{2V}\int\d{\vec r} & \beta W(\vec r)
   \left(\rho_+(\vec r)+\rho_-(\vec r)\right) = \\
=& \frac{1}{2V} \frac{1}{(2\pi)^3}
   \frac{\bar\rho_+-\bar\rho_-}{\bar\rho_++\bar\rho_-}
   \int \d{\vec k} \beta W_{\vec k}\rho_{-{\vec k}}\\
=& \frac{1}{V} \frac{A}{2}
   \left(\frac{Ze^{\bar\kappa a}}{1+\bar\kappa a}\right)^2
   \sum_{i<j}\frac{\lB e^{-\bar\kappa R_{ij}}}{R_{ij}},
\end{split}
\end{align}
where the Fourier transform of $W(\vec r)$ is given by
Eq.~\eqref{eq:fourierW} and where we used that $W(\vec
r)[\rho_+(\vec r)/\bar\rho_+ + \rho_-(\vec r)/\bar\rho_-]\equiv0$.  

Substitution of Eqs.~\eqref{eq:electroint}
and~\eqref{eq:hardcoreint} into the grand
potential~\eqref{eq:grandpot_gen} leads to
\begin{align}\label{eq:grandpot2}
\begin{split}
\beta\Omega = &
    (1+A)\left(\frac{Ze^{\bar\kappa a}}{1+\bar\kappa a}\right)^2
        \sum_{i<j} \lB\frac{e^{-\bar\kappa R_{ij}}}{R_{ij}} \\
 &  -Z^2\sum_{i<j}\frac{\lB}{R_{ij}}
    + \beta\Phi, \\
\end{split}
\end{align}
where the ``volume term'' $\Phi=\PhiD+\Phi_0$ is given by
Eqs.~\eqref{eq:donnanterm} and~\eqref{eq:volterm1}.

Gathering Eqs.~\eqref{H}, \eqref{Vcc} and~\eqref{eq:grandpot2}, we find that
the effective Hamiltonian can be cast into the form given by
Eqs.~\eqref{H2}--\eqref{V2}.

%%%%%%%%%%%%%%%%%%%%%%%%%%%%%%%%%%%%%%%%%%%%%%%%%%%%%%%%%%%%%
\section{Alternative hard-core terms}\label{appB}
\noindent 
We have already mentioned that, in this paper, we use a
slightly different definition of the hard-core parameters $\beta
v_0$ and $\beta w_0$ than were used by Van Roij and Hansen in
Refs.~\cite{PRL-79-3082,PRE-59-2010}.  In this Appendix we
make this statement explicit and calculate, within the
grand-canonical scheme of this work, the effective Hamiltonian
using the definition of the hard-core potentials of
Refs.~\cite{PRL-79-3082,PRE-59-2010}.

In contrast to the presently used definition~$U_\pm(\vec r)=\pm
V(\vec r)+W(\vec r)$ for the micro-ion--colloid interactions, as
outlined just above Eq.~\eqref{v}, Van Roij and Hansen used
the definition
\begin{align}
U_\pm(\vec r) = \pm V(\vec r) +
\frac{2\bar\rho_\mp}{\bar\rho_++\bar\rho_-} W(\vec r),
\end{align}
where the potentials~$V(\vec r)$ and~$W(\vec r)$ are defined in
Eqs.~\eqref{v} and~\eqref{w}.  With this definition, the grand
potential becomes
\begin{align}
\nonumber
\Omega[\rho_+,\rho&_-] =
 \Omega'_{\rm id}[\rho_+]
 +\Omega'_{\rm id}[\rho_-]
 + \int \d{\vec r} \rho({\vec r})V({\vec r}) \\
\label{B:functional}
& +\kB T\lB\int \d{\vec r}\,\d{\vec r'}
     \frac{\rho({\vec r})\rho({\vec r}')}{|{\vec r}-{\vec r}'|}
     \\
\nonumber
&+ \frac{2\bar\rho_+\bar\rho_-}{\bar\rho_++\bar\rho_-}
   \int \d{\vec r}
   \left(\frac{\rho_+(\vec r)}{\bar\rho_+}
   + \frac{\rho_-(\vec r)}{\bar\rho_-}\right)
   W({\vec r}),
\end{align}
where the ideal-gas functionals~$\beta\Omega_{\rm id}[\rho_\pm]$ are
defined in Eq.~\eqref{Oidapprox}.

The corresponding Euler--Lagrange equations are then given by
\begin{multline}
\ln\frac{\bar\rho_\pm}{{\cs}}
+\frac{\rho_\pm(\vec r)
-\bar\rho_\pm}{\bar\rho_\pm}
\pm \phi(\vec r)
+ \frac{2\bar\rho_\mp \beta W(\vec r)}{\bar\rho_++\bar\rho_-}=0.
\label{B:el1}
\end{multline}
By integrating these equations over the system volume, and using
the condition for global charge neutrality, we find that the
average densities~$\bar\rho_\pm$ are identical to those given in
Eq.~\eqref{eq:salt}. The Donnan potential~$\bar\phi$, however, is
not given by Eq.~\eqref{donnan} anymore, but by
\begin{align}
\bar\phi =
-\sinh^{-1}\left[\frac{Zn}{2\cs} e^{\eta\beta w_0}\right]
+ \eta\beta w_0 \frac{Zn}{\bar\rho_++\bar\rho_-}
\end{align}
instead. Although this expression also reduces to the usual Donnan
expression in the limit $n\rightarrow0$, it is physically less
satisfactory than the result we found in Eq.~\eqref{donnan}, as at
high~$\eta$ its sign can become different from that of the colloidal
charge~$-Ze$.

\medskip

To calculate the density profiles, we take the following linear
combination of the Euler--Lagrange equations~\eqref{B:el1}:
\begin{subequations}
\begin{align}
\label{B:el2a}
  \frac{\rho_+(\vec r)}{\bar\rho_+}
  + \frac{\rho_-(\vec r)}{\bar\rho_-}
  =& 2\left(1-\beta W(\vec r)+\eta\beta w_0\right); \\
\label{B:el2b}
  \frac{\rho(\vec r) - \bar\rho}{\bar\rho_++\bar\rho_-} =&
  -(\phi(\vec r)-\bar\phi).
\end{align}
\end{subequations}
Note that, due to the different definition of~$U_\pm(\vec r)$, the
hard-core potential~$W(\vec r)$ is now totally decoupled from the
charge density~$\rho(\vec r)$.

Eq.~\eqref{B:el2a} is identical to Eq.~\eqref{el2a}, so its
solution is again $0$ inside the hard cores of the colloids, and
given by Eq.~\eqref{el2a2} outside the hard cores, provided that we
fix $\beta w_0=1/(1-\eta)$.

The second equation~\eqref{B:el2b} is {\em not} identical to its
counterpart Eq.~\eqref{el2b}.  The solution is quite similar though:
we need to fix the hard core parameter~$\beta v_0$ to
\begin{align}
\beta v_0 = -Z\frac{\bar\kappa\lB}{1+\bar\kappa a},
\end{align}
in order to make sure that the charge density~$\rho(\vec r)$ is a
multi-centered sum of DLVO profiles.  The solution (in $\vec
k$-space) is then given by
\begin{align}\label{B:rho_sol}
\begin{split}
\fourier\rho =
 & (2\pi)^3 \left\{\bar\rho+(\bar\rho_++\bar\rho_-)\bar\phi\right\}
   \frac{k^2}{k^2+\bar\kappa^2} \delta(\vec k) \\
 & +\frac{Z}{1+\bar\kappa a}
    \frac{\cos ka+\frac{\bar\kappa}{k}\sin ka}
         {1+k^2/\bar\kappa^2}
    \sum_j e^{i\vec k\cdot\vec R_j},
\end{split}
\end{align}
which, in real space, is indeed a multi-centered sum $\rho(\vec
r)=\sum_i\rho_1(|\vec r-\vec R_i|)$ with the individual profiles
given by Eq.~\eqref{eq:charge1}.  Note that Eq.~\eqref{eq:rho_sol}
and~\eqref{B:rho_sol} only differ in the $\propto k^2\delta(\vec
k)$ term.  As a consequence, the profiles~$\rho(\vec r)$ resulting
from those two equations, are identical, but the minimum of the
functional differs.

\medskip

Upon insertion of the equilibrium density profiles into the
functional~\eqref{B:functional}, we immediately notice that the last
term on the right hand side vanishes.  The grand potential then
becomes
\begin{align}
\label{B:minumum}
\begin{split}
\frac{\beta\Omega}{V} =
&
  \sum_\pm \bar\rho_\pm\left(\ln\left[\frac{\bar\rho_\pm}{{\cs}}\right]-1\right)
  + \frac{Zn}{2}\bar\phi  \\
& + \frac{\eta}{1-\eta}
    \frac{2\bar\rho_+\bar\rho_-}{\bar\rho_+ + \bar\rho_-}
  + \frac{1}{2V}\int\d{\vec r}\rho(\vec r)\beta V(\vec r).
\end{split}
\end{align}
The integral in this expression can now be calculated using
Parseval's theorem and the expression~\eqref{eq:fourierV} for the
Fourier transform of~$V(\vec r)$.  The result is
\begin{align}\begin{split}
\label{B:electroint}
\frac{1}{2V} & \int\d{\vec r}\rho(\vec r)\beta V(\vec r)
= \\
&   \frac{1}{V} \sum_{i<j}\left\{
    \left(\frac{Ze^{\bar\kappa a}}{1+\bar\kappa a}\right)^2
    \frac{\lB e^{-\bar\kappa R_{ij}}}{R_{ij}}
    - Z^2 \frac{\lB}{R_{ij}} \right\} \\
& - \frac{\np}{2}\frac{Z^2\bar\kappa\lB}{1+\bar\kappa a}
  - \frac{1}{2}\frac{(\bar\rho_+-\bar\rho_-)^2}{\bar\rho_++\bar\rho_-}
  - \frac{Z\np}{2}\bar\phi, \\
\end{split}\end{align}
so that the grand potential eventually becomes
\begin{align}\label{B:grandpot2}
\begin{split}
\beta\Omega = &
    \left(\frac{Ze^{\bar\kappa a}}{1+\bar\kappa a}\right)^2
        \sum_{i<j} \lB\frac{e^{-\bar\kappa R_{ij}}}{R_{ij}} \\
 &  -Z^2\sum_{i<j}\frac{\lB}{R_{ij}}
    + \beta\Phi.
\end{split}
\end{align}
The volume term~$\beta\Phi$ is exactly equal to the one that was
found previously in Eqs.~\eqref{eq:donnanterm}
and~\eqref{eq:volterm1};  the colloidal pair interaction,
however, reduces to a purely DLVO interaction, i.e.~the factor~$A$
we found before, is now equal to~$0$.

\bibliographystyle{myapsrev}
%\bibliography{colloids}

\begin{thebibliography}{100}
\expandafter\ifx\csname natexlab\endcsname\relax\def\natexlab#1{#1}\fi
\expandafter\ifx\csname bibnamefont\endcsname\relax
  \def\bibnamefont#1{#1}\fi
\expandafter\ifx\csname bibfnamefont\endcsname\relax
  \def\bibfnamefont#1{#1}\fi
\expandafter\ifx\csname citenamefont\endcsname\relax
  \def\citenamefont#1{#1}\fi
\expandafter\ifx\csname url\endcsname\relax
  \def\url#1{\texttt{#1}}\fi
\expandafter\ifx\csname urlprefix\endcsname\relax\def\urlprefix{}\fi
\providecommand{\bibinfo}[2]{#2}
\providecommand{\eprint}[2][]{\url{#2}}

\bibitem[{\citenamefont{Asakura and Osawa}(1954)}]{AO-depl}
\bibinfo{author}{\bibfnamefont{S.}~\bibnamefont{Asakura}} \bibnamefont{and}
  \bibinfo{author}{\bibfnamefont{F.}~\bibnamefont{Osawa}}, \bibinfo{journal}{J.
  Chem. Phys.} \textbf{\bibinfo{volume}{22}}, \bibinfo{pages}{1255}
  (\bibinfo{year}{1954}).

\bibitem[{\citenamefont{Vrij}(1976)}]{vrij-depl}
\bibinfo{author}{\bibfnamefont{A.}~\bibnamefont{Vrij}}, \bibinfo{journal}{Pure
  Appl.\ Chem.} \textbf{\bibinfo{volume}{48}}, \bibinfo{pages}{471}
  (\bibinfo{year}{1976}).

\bibitem[{\citenamefont{Levin}(2002)}]{levin-review}
\bibinfo{author}{\bibfnamefont{Y.}~\bibnamefont{Levin}}, \bibinfo{journal}{Rep.
  Prog. Phys.} \textbf{\bibinfo{volume}{65}}, \bibinfo{pages}{1577}
  (\bibinfo{year}{2002}).

\bibitem[{\citenamefont{Hansen and L{\"o}wen}(2000)}]{hansen-lowen-review}
\bibinfo{author}{\bibfnamefont{J.-P.} \bibnamefont{Hansen}} \bibnamefont{and}
  \bibinfo{author}{\bibfnamefont{H.}~\bibnamefont{L{\"o}wen}},
  \bibinfo{journal}{Annu. Rev. Phys. Chem.} \textbf{\bibinfo{volume}{51}},
  \bibinfo{pages}{209} (\bibinfo{year}{2000}).

\bibitem[{\citenamefont{Likos}(2001)}]{likos-review}
\bibinfo{author}{\bibfnamefont{C.~N.} \bibnamefont{Likos}},
  \bibinfo{journal}{Phys. Rep.} \textbf{\bibinfo{volume}{348}},
  \bibinfo{pages}{267} (\bibinfo{year}{2001}).

\bibitem[{\citenamefont{Dijkstra and van Roij}(2002)}]{PRL-89-208303}
\bibinfo{author}{\bibfnamefont{M.}~\bibnamefont{Dijkstra}} \bibnamefont{and}
  \bibinfo{author}{\bibfnamefont{R.}~\bibnamefont{van Roij}},
  \bibinfo{journal}{Phys. Rev. Lett.} \textbf{\bibinfo{volume}{89}},
  \bibinfo{pages}{208303} (\bibinfo{year}{2002}).

\bibitem[{\citenamefont{Brader et~al.}(2001)\citenamefont{Brader, Evans,
  Schmidt, and L{\"o}wen}}]{JPCM-14-L1}
\bibinfo{author}{\bibfnamefont{J.~M.} \bibnamefont{Brader}},
  \bibinfo{author}{\bibfnamefont{R.}~\bibnamefont{Evans}},
  \bibinfo{author}{\bibfnamefont{M.}~\bibnamefont{Schmidt}}, \bibnamefont{and}
  \bibinfo{author}{\bibfnamefont{H.}~\bibnamefont{L{\"o}wen}},
  \bibinfo{journal}{J. Phys.: Condens. Matter} \textbf{\bibinfo{volume}{14}},
  \bibinfo{pages}{L1} (\bibinfo{year}{2001}).

\bibitem[{\citenamefont{Russ et~al.}(2002)\citenamefont{Russ, von Gr{\"
  u}nberg, Dijkstra, and van Roij}}]{PRE-66-011402}
\bibinfo{author}{\bibfnamefont{C.}~\bibnamefont{Russ}},
  \bibinfo{author}{\bibfnamefont{H.~H.} \bibnamefont{von Gr{\" u}nberg}},
  \bibinfo{author}{\bibfnamefont{M.}~\bibnamefont{Dijkstra}}, \bibnamefont{and}
  \bibinfo{author}{\bibfnamefont{R.}~\bibnamefont{van Roij}},
  \bibinfo{journal}{Phys. Rev. E} \textbf{\bibinfo{volume}{66}},
  \bibinfo{pages}{011402} (\bibinfo{year}{2002}).

\bibitem[{\citenamefont{Derjaguin and Landau}(1941)}]{dl1941}
\bibinfo{author}{\bibfnamefont{B.}~\bibnamefont{Derjaguin}} \bibnamefont{and}
  \bibinfo{author}{\bibfnamefont{L.}~\bibnamefont{Landau}},
  \bibinfo{journal}{Acta Physicochim., URSS} \textbf{\bibinfo{volume}{14}},
  \bibinfo{pages}{633} (\bibinfo{year}{1941}).

\bibitem[{\citenamefont{Verwey and Overbeek}(1948)}]{vo1948}
\bibinfo{author}{\bibfnamefont{J.~W.} \bibnamefont{Verwey}} \bibnamefont{and}
  \bibinfo{author}{\bibfnamefont{J.~T.~G.} \bibnamefont{Overbeek}},
  \emph{\bibinfo{title}{Theory of the stability of lyotropic colloids}}
  (\bibinfo{publisher}{Elsevier, Amsterdam}, \bibinfo{year}{1948}).

\bibitem[{\citenamefont{Alder and Wainwright}(1957)}]{JCP-27-1208}
\bibinfo{author}{\bibfnamefont{B.~J.} \bibnamefont{Alder}} \bibnamefont{and}
  \bibinfo{author}{\bibfnamefont{T.~E.} \bibnamefont{Wainwright}},
  \bibinfo{journal}{J. Chem. Phys.} \textbf{\bibinfo{volume}{27}},
  \bibinfo{pages}{1208} (\bibinfo{year}{1957}).

\bibitem[{\citenamefont{Pusey et~al.}(1989)\citenamefont{Pusey, van Megen,
  Bartlett, Ackerson, Rarity, and Underwood}}]{PRL-63-2753}
\bibinfo{author}{\bibfnamefont{P.~N.} \bibnamefont{Pusey}},
  \bibinfo{author}{\bibfnamefont{W.}~\bibnamefont{van Megen}},
  \bibinfo{author}{\bibfnamefont{P.}~\bibnamefont{Bartlett}},
  \bibinfo{author}{\bibfnamefont{B.~J.} \bibnamefont{Ackerson}},
  \bibinfo{author}{\bibfnamefont{J.~G.} \bibnamefont{Rarity}},
  \bibnamefont{and} \bibinfo{author}{\bibfnamefont{S.~M.}
  \bibnamefont{Underwood}}, \bibinfo{journal}{Phys. Rev. Lett.}
  \textbf{\bibinfo{volume}{63}}, \bibinfo{pages}{2753} (\bibinfo{year}{1989}).

\bibitem[{\citenamefont{Stevens and Robbins}(1990)}]{EPL-12-81}
\bibinfo{author}{\bibfnamefont{M.~J.} \bibnamefont{Stevens}} \bibnamefont{and}
  \bibinfo{author}{\bibfnamefont{M.~O.} \bibnamefont{Robbins}},
  \bibinfo{journal}{Europhys. Lett.} \textbf{\bibinfo{volume}{12}},
  \bibinfo{pages}{81} (\bibinfo{year}{1990}).

\bibitem[{\citenamefont{Hynninen and
  Dijkstra}(2005{\natexlab{a}})}]{PRL-94-138303}
\bibinfo{author}{\bibfnamefont{A.-P.} \bibnamefont{Hynninen}} \bibnamefont{and}
  \bibinfo{author}{\bibfnamefont{M.}~\bibnamefont{Dijkstra}},
  \bibinfo{journal}{Phys. Rev. Lett.} \textbf{\bibinfo{volume}{94}},
  \bibinfo{pages}{138303} (\bibinfo{year}{2005}{\natexlab{a}}).

\bibitem[{\citenamefont{Reus et~al.}(1997)\citenamefont{Reus, Belloni, Zemb,
  Lutterbach, and Vesmold}}]{reus1997}
\bibinfo{author}{\bibfnamefont{V.}~\bibnamefont{Reus}},
  \bibinfo{author}{\bibfnamefont{L.}~\bibnamefont{Belloni}},
  \bibinfo{author}{\bibfnamefont{T.}~\bibnamefont{Zemb}},
  \bibinfo{author}{\bibfnamefont{N.}~\bibnamefont{Lutterbach}},
  \bibnamefont{and} \bibinfo{author}{\bibfnamefont{H.}~\bibnamefont{Vesmold}},
  \bibinfo{journal}{J. Phys. II France} \textbf{\bibinfo{volume}{7}},
  \bibinfo{pages}{603} (\bibinfo{year}{1997}).

\bibitem[{\citenamefont{Reus et~al.}(1999)\citenamefont{Reus, Belloni, Zemb,
  Lutterbach, and Versmold}}]{reus1999}
\bibinfo{author}{\bibfnamefont{V.}~\bibnamefont{Reus}},
  \bibinfo{author}{\bibfnamefont{L.}~\bibnamefont{Belloni}},
  \bibinfo{author}{\bibfnamefont{T.}~\bibnamefont{Zemb}},
  \bibinfo{author}{\bibfnamefont{N.}~\bibnamefont{Lutterbach}},
  \bibnamefont{and} \bibinfo{author}{\bibfnamefont{H.}~\bibnamefont{Versmold}},
  \bibinfo{journal}{Colloids Surf., A} \textbf{\bibinfo{volume}{151}},
  \bibinfo{pages}{449} (\bibinfo{year}{1999}).

\bibitem[{\citenamefont{Sirota et~al.}(1989)\citenamefont{Sirota, Ou-Yang,
  Sinha, Chaikin, Axe, and Fujii}}]{PRL-62-1524}
\bibinfo{author}{\bibfnamefont{E.~B.} \bibnamefont{Sirota}},
  \bibinfo{author}{\bibfnamefont{H.~D.} \bibnamefont{Ou-Yang}},
  \bibinfo{author}{\bibfnamefont{S.~K.} \bibnamefont{Sinha}},
  \bibinfo{author}{\bibfnamefont{P.~M.} \bibnamefont{Chaikin}},
  \bibinfo{author}{\bibfnamefont{J.~D.} \bibnamefont{Axe}}, \bibnamefont{and}
  \bibinfo{author}{\bibfnamefont{Y.}~\bibnamefont{Fujii}},
  \bibinfo{journal}{Phys. Rev. Lett.} \textbf{\bibinfo{volume}{62}},
  \bibinfo{pages}{1524} (\bibinfo{year}{1989}).

\bibitem[{\citenamefont{Chen et~al.}(1988)\citenamefont{Chen, Sheu, Kalus, and
  Hoffman}}]{scatter1}
\bibinfo{author}{\bibfnamefont{S.~H.} \bibnamefont{Chen}},
  \bibinfo{author}{\bibfnamefont{E.~Y.} \bibnamefont{Sheu}},
  \bibinfo{author}{\bibfnamefont{J.}~\bibnamefont{Kalus}}, \bibnamefont{and}
  \bibinfo{author}{\bibfnamefont{H.}~\bibnamefont{Hoffman}},
  \bibinfo{journal}{J. Applied Cryst.} \textbf{\bibinfo{volume}{21}},
  \bibinfo{pages}{751} (\bibinfo{year}{1988}).

\bibitem[{\citenamefont{Royall et~al.}(2003)\citenamefont{Royall, Leunissen,
  and van Blaaderen}}]{paddy2003}
\bibinfo{author}{\bibfnamefont{C.~P.} \bibnamefont{Royall}},
  \bibinfo{author}{\bibfnamefont{M.~E.} \bibnamefont{Leunissen}},
  \bibnamefont{and} \bibinfo{author}{\bibfnamefont{A.}~\bibnamefont{van
  Blaaderen}}, \bibinfo{journal}{J. Phys.: Condens. Matter}
  \textbf{\bibinfo{volume}{15}}, \bibinfo{pages}{S3581} (\bibinfo{year}{2003}).

\bibitem[{\citenamefont{Crocker and Grier}(1994)}]{PRL-73-352}
\bibinfo{author}{\bibfnamefont{J.~C.} \bibnamefont{Crocker}} \bibnamefont{and}
  \bibinfo{author}{\bibfnamefont{D.~G.} \bibnamefont{Grier}},
  \bibinfo{journal}{Phys. Rev. Lett.} \textbf{\bibinfo{volume}{73}},
  \bibinfo{pages}{352} (\bibinfo{year}{1994}).

\bibitem[{\citenamefont{Vondermassen et~al.}(1994)\citenamefont{Vondermassen,
  Bongers, Mueller, and Versmold}}]{Langmuir-10-1351}
\bibinfo{author}{\bibfnamefont{K.}~\bibnamefont{Vondermassen}},
  \bibinfo{author}{\bibfnamefont{J.}~\bibnamefont{Bongers}},
  \bibinfo{author}{\bibfnamefont{A.}~\bibnamefont{Mueller}}, \bibnamefont{and}
  \bibinfo{author}{\bibfnamefont{H.}~\bibnamefont{Versmold}},
  \bibinfo{journal}{Langmuir} \textbf{\bibinfo{volume}{10}},
  \bibinfo{pages}{1351} (\bibinfo{year}{1994}).

\bibitem[{\citenamefont{Ito et~al.}(1994)\citenamefont{Ito, Hiroshi, and
  Ise}}]{ise1994}
\bibinfo{author}{\bibfnamefont{K.}~\bibnamefont{Ito}},
  \bibinfo{author}{\bibfnamefont{H.}~\bibnamefont{Hiroshi}}, \bibnamefont{and}
  \bibinfo{author}{\bibfnamefont{N.}~\bibnamefont{Ise}},
  \bibinfo{journal}{Science} \textbf{\bibinfo{volume}{263}},
  \bibinfo{pages}{66} (\bibinfo{year}{1994}).

\bibitem[{\citenamefont{Tata et~al.}(1997)\citenamefont{Tata, Yamahara,
  Rajamani, and Ise}}]{PRL-78-2660}
\bibinfo{author}{\bibfnamefont{B.~V.~R.} \bibnamefont{Tata}},
  \bibinfo{author}{\bibfnamefont{E.}~\bibnamefont{Yamahara}},
  \bibinfo{author}{\bibfnamefont{P.~V.} \bibnamefont{Rajamani}},
  \bibnamefont{and} \bibinfo{author}{\bibfnamefont{N.}~\bibnamefont{Ise}},
  \bibinfo{journal}{Phys. Rev. Lett.} \textbf{\bibinfo{volume}{78}},
  \bibinfo{pages}{2660} (\bibinfo{year}{1997}).

\bibitem[{\citenamefont{Ise et~al.}(1983)\citenamefont{Ise, Okubo, Sugimura,
  Ito, and Nolte}}]{ise1983}
\bibinfo{author}{\bibfnamefont{N.}~\bibnamefont{Ise}},
  \bibinfo{author}{\bibfnamefont{T.}~\bibnamefont{Okubo}},
  \bibinfo{author}{\bibfnamefont{M.}~\bibnamefont{Sugimura}},
  \bibinfo{author}{\bibfnamefont{K.}~\bibnamefont{Ito}}, \bibnamefont{and}
  \bibinfo{author}{\bibfnamefont{H.}~\bibnamefont{Nolte}}, \bibinfo{journal}{J.
  Chem. Phys.} \textbf{\bibinfo{volume}{78}}, \bibinfo{pages}{536}
  (\bibinfo{year}{1983}).

\bibitem[{\citenamefont{Larsen and Grier}(1997)}]{Nature-385-230}
\bibinfo{author}{\bibfnamefont{A.~E.} \bibnamefont{Larsen}} \bibnamefont{and}
  \bibinfo{author}{\bibfnamefont{D.~G.} \bibnamefont{Grier}},
  \bibinfo{journal}{Nature} \textbf{\bibinfo{volume}{385}},
  \bibinfo{pages}{230} (\bibinfo{year}{1997}).

\bibitem[{\citenamefont{Tata et~al.}(1992)\citenamefont{Tata, Rajalakshmi, and
  Arora}}]{tataarora}
\bibinfo{author}{\bibfnamefont{B.~V.~R.} \bibnamefont{Tata}},
  \bibinfo{author}{\bibfnamefont{M.}~\bibnamefont{Rajalakshmi}},
  \bibnamefont{and} \bibinfo{author}{\bibfnamefont{A.~K.} \bibnamefont{Arora}},
  \bibinfo{journal}{Phys. Rev. Lett.} \textbf{\bibinfo{volume}{69}},
  \bibinfo{pages}{3778} (\bibinfo{year}{1992}).

\bibitem[{\citenamefont{Palberg and W{\"u}rth}(1994)}]{palberg1994}
\bibinfo{author}{\bibfnamefont{T.}~\bibnamefont{Palberg}} \bibnamefont{and}
  \bibinfo{author}{\bibfnamefont{M.}~\bibnamefont{W{\"u}rth}},
  \bibinfo{journal}{Phys. Rev. Lett.} \textbf{\bibinfo{volume}{72}},
  \bibinfo{pages}{786} (\bibinfo{year}{1994}).

\bibitem[{\citenamefont{Tata and Arora}(1994)}]{arora_reply}
\bibinfo{author}{\bibfnamefont{B.~V.~R.} \bibnamefont{Tata}} \bibnamefont{and}
  \bibinfo{author}{\bibfnamefont{A.~K.} \bibnamefont{Arora}},
  \bibinfo{journal}{Phys. Rev. Lett.} \textbf{\bibinfo{volume}{72}},
  \bibinfo{pages}{787} (\bibinfo{year}{1994}).

\bibitem[{\citenamefont{Hynninen et~al.}(2003)\citenamefont{Hynninen, Dijkstra,
  and Roij}}]{JPCM-15-S3549}
\bibinfo{author}{\bibfnamefont{A.-P.} \bibnamefont{Hynninen}},
  \bibinfo{author}{\bibfnamefont{M.}~\bibnamefont{Dijkstra}}, \bibnamefont{and}
  \bibinfo{author}{\bibfnamefont{R.~v.} \bibnamefont{Roij}},
  \bibinfo{journal}{J. Phys.: Condens. Matter} \textbf{\bibinfo{volume}{15}},
  \bibinfo{pages}{S3549} (\bibinfo{year}{2003}).

\bibitem[{\citenamefont{Hynninen et~al.}(2004)\citenamefont{Hynninen, Dijkstra,
  and van Roij}}]{PRE-69-061407}
\bibinfo{author}{\bibfnamefont{A.-P.} \bibnamefont{Hynninen}},
  \bibinfo{author}{\bibfnamefont{M.}~\bibnamefont{Dijkstra}}, \bibnamefont{and}
  \bibinfo{author}{\bibfnamefont{R.}~\bibnamefont{van Roij}},
  \bibinfo{journal}{Phys. Rev. E} \textbf{\bibinfo{volume}{69}},
  \bibinfo{pages}{061407} (\bibinfo{year}{2004}).

\bibitem[{\citenamefont{Russ et~al.}(2005)\citenamefont{Russ, Brunner,
  Bechinger, and von Gr{\" u}nberg}}]{EPL-69-468}
\bibinfo{author}{\bibfnamefont{C.}~\bibnamefont{Russ}},
  \bibinfo{author}{\bibfnamefont{M.}~\bibnamefont{Brunner}},
  \bibinfo{author}{\bibfnamefont{C.}~\bibnamefont{Bechinger}},
  \bibnamefont{and} \bibinfo{author}{\bibfnamefont{H.~H.} \bibnamefont{von
  Gr{\" u}nberg}}, \bibinfo{journal}{Europhys. Lett.}
  \textbf{\bibinfo{volume}{69}}, \bibinfo{pages}{468} (\bibinfo{year}{2005}).

\bibitem[{\citenamefont{Russ et~al.}(2003)\citenamefont{Russ, Zahn, and von
  Gr{\"u}nberg}}]{JPCM-15-S3509}
\bibinfo{author}{\bibfnamefont{C.}~\bibnamefont{Russ}},
  \bibinfo{author}{\bibfnamefont{K.}~\bibnamefont{Zahn}}, \bibnamefont{and}
  \bibinfo{author}{\bibfnamefont{H.-H.} \bibnamefont{von Gr{\"u}nberg}},
  \bibinfo{journal}{J. Phys.: Condens. Matter} \textbf{\bibinfo{volume}{15}},
  \bibinfo{pages}{S3509} (\bibinfo{year}{2003}).

\bibitem[{\citenamefont{Zahn et~al.}(2003)\citenamefont{Zahn, Maret, Russ, and
  von Gr{\"u}nberg}}]{PRL-91-115502}
\bibinfo{author}{\bibfnamefont{K.}~\bibnamefont{Zahn}},
  \bibinfo{author}{\bibfnamefont{G.}~\bibnamefont{Maret}},
  \bibinfo{author}{\bibfnamefont{C.}~\bibnamefont{Russ}}, \bibnamefont{and}
  \bibinfo{author}{\bibfnamefont{H.~H.} \bibnamefont{von Gr{\"u}nberg}},
  \bibinfo{journal}{Phys. Rev. Lett.} \textbf{\bibinfo{volume}{91}},
  \bibinfo{pages}{115502} (\bibinfo{year}{2003}).

\bibitem[{\citenamefont{Brunner et~al.}(2004)\citenamefont{Brunner, Dobnikar,
  von Grunberg, and Bechinger}}]{PRL-92-078301}
\bibinfo{author}{\bibfnamefont{M.}~\bibnamefont{Brunner}},
  \bibinfo{author}{\bibfnamefont{J.}~\bibnamefont{Dobnikar}},
  \bibinfo{author}{\bibfnamefont{H.~H.} \bibnamefont{von Grunberg}},
  \bibnamefont{and}
  \bibinfo{author}{\bibfnamefont{C.}~\bibnamefont{Bechinger}},
  \bibinfo{journal}{Phys. Rev. Lett.} \textbf{\bibinfo{volume}{92}},
  \bibinfo{pages}{078301} (\bibinfo{year}{2004}).

\bibitem[{\citenamefont{Dobnikar et~al.}(2004)\citenamefont{Dobnikar, Brunner,
  von Grunberg, and Bechinger}}]{PRE-69-031402}
\bibinfo{author}{\bibfnamefont{J.}~\bibnamefont{Dobnikar}},
  \bibinfo{author}{\bibfnamefont{M.}~\bibnamefont{Brunner}},
  \bibinfo{author}{\bibfnamefont{H.~H.} \bibnamefont{von Grunberg}},
  \bibnamefont{and}
  \bibinfo{author}{\bibfnamefont{C.}~\bibnamefont{Bechinger}},
  \bibinfo{journal}{Phys. Rev. E} \textbf{\bibinfo{volume}{69}},
  \bibinfo{pages}{031402} (\bibinfo{year}{2004}).

\bibitem[{\citenamefont{Hynninen and
  Dijkstra}(2005{\natexlab{b}})}]{JCP-123-244902}
\bibinfo{author}{\bibfnamefont{A.-P.} \bibnamefont{Hynninen}} \bibnamefont{and}
  \bibinfo{author}{\bibfnamefont{M.}~\bibnamefont{Dijkstra}},
  \bibinfo{journal}{J. Chem. Phys.} \textbf{\bibinfo{volume}{123}},
  \bibinfo{eid}{244902} (\bibinfo{year}{2005}{\natexlab{b}}).

\bibitem[{\citenamefont{Dijkstra and van Roij}(1998)}]{JPCM-10-1219}
\bibinfo{author}{\bibfnamefont{M.}~\bibnamefont{Dijkstra}} \bibnamefont{and}
  \bibinfo{author}{\bibfnamefont{R.}~\bibnamefont{van Roij}},
  \bibinfo{journal}{J. Phys.: Condens. Matter} \textbf{\bibinfo{volume}{10}},
  \bibinfo{pages}{1219} (\bibinfo{year}{1998}).

\bibitem[{\citenamefont{Warren}(2000{\natexlab{a}})}]{JCP-112-4683}
\bibinfo{author}{\bibfnamefont{P.~B.} \bibnamefont{Warren}},
  \bibinfo{journal}{J. Chem. Phys.} \textbf{\bibinfo{volume}{112}},
  \bibinfo{pages}{4683} (\bibinfo{year}{2000}{\natexlab{a}}).

\bibitem[{\citenamefont{Diehl et~al.}(2001)\citenamefont{Diehl, Barbosa, and
  Levin}}]{EPL-53-86}
\bibinfo{author}{\bibfnamefont{A.}~\bibnamefont{Diehl}},
  \bibinfo{author}{\bibfnamefont{M.~C.} \bibnamefont{Barbosa}},
  \bibnamefont{and} \bibinfo{author}{\bibfnamefont{Y.}~\bibnamefont{Levin}},
  \bibinfo{journal}{Europhys. Lett.} \textbf{\bibinfo{volume}{53}},
  \bibinfo{pages}{86} (\bibinfo{year}{2001}).

\bibitem[{\citenamefont{Warren}(2000{\natexlab{b}})}]{warren-condmat-2000}
\bibinfo{author}{\bibfnamefont{P.~B.} \bibnamefont{Warren}}
  (\bibinfo{year}{2000}{\natexlab{b}}), \urlprefix\url{cond-mat/0006289}.

\bibitem[{\citenamefont{Denton}(2000)}]{PRE-62-3855}
\bibinfo{author}{\bibfnamefont{A.~R.} \bibnamefont{Denton}},
  \bibinfo{journal}{Phys. Rev. E} \textbf{\bibinfo{volume}{62}},
  \bibinfo{pages}{3855} (\bibinfo{year}{2000}).

\bibitem[{\citenamefont{Petris and Chan}(2002)}]{JCP-116-8588}
\bibinfo{author}{\bibfnamefont{S.~N.} \bibnamefont{Petris}} \bibnamefont{and}
  \bibinfo{author}{\bibfnamefont{D.~Y.~C.} \bibnamefont{Chan}},
  \bibinfo{journal}{J. Chem. Phys.} \textbf{\bibinfo{volume}{116}},
  \bibinfo{pages}{8588} (\bibinfo{year}{2002}).

\bibitem[{\citenamefont{Chan et~al.}(2001)\citenamefont{Chan, Linse, and
  Petris}}]{Lang-17-4202}
\bibinfo{author}{\bibfnamefont{D.~Y.~C.} \bibnamefont{Chan}},
  \bibinfo{author}{\bibfnamefont{P.}~\bibnamefont{Linse}}, \bibnamefont{and}
  \bibinfo{author}{\bibfnamefont{S.~N.} \bibnamefont{Petris}},
  \bibinfo{journal}{Langmuir} \textbf{\bibinfo{volume}{17}},
  \bibinfo{pages}{4202} (\bibinfo{year}{2001}).

\bibitem[{\citenamefont{Warren}(2003)}]{JPCM-15-S3467}
\bibinfo{author}{\bibfnamefont{P.~B.} \bibnamefont{Warren}},
  \bibinfo{journal}{J. Phys.: Condens. Matter} \textbf{\bibinfo{volume}{15}},
  \bibinfo{pages}{S3467} (\bibinfo{year}{2003}).

\bibitem[{\citenamefont{Beresford-Smith
  et~al.}(1985)\citenamefont{Beresford-Smith, Chan, and
  Mitchell}}]{JCIS-105-216}
\bibinfo{author}{\bibfnamefont{B.}~\bibnamefont{Beresford-Smith}},
  \bibinfo{author}{\bibfnamefont{D.~Y.~C.} \bibnamefont{Chan}},
  \bibnamefont{and} \bibinfo{author}{\bibfnamefont{D.~J.}
  \bibnamefont{Mitchell}}, \bibinfo{journal}{J. Colloid Interface Sci.}
  \textbf{\bibinfo{volume}{105}}, \bibinfo{pages}{216} (\bibinfo{year}{1985}).

\bibitem[{\citenamefont{van Roij and Hansen}(1997)}]{PRL-79-3082}
\bibinfo{author}{\bibfnamefont{R.}~\bibnamefont{van Roij}} \bibnamefont{and}
  \bibinfo{author}{\bibfnamefont{J.-P.} \bibnamefont{Hansen}},
  \bibinfo{journal}{Phys. Rev. Lett.} \textbf{\bibinfo{volume}{79}},
  \bibinfo{pages}{3082} (\bibinfo{year}{1997}).

\bibitem[{\citenamefont{van Roij et~al.}(1999)\citenamefont{van Roij, Dijkstra,
  and Hansen}}]{PRE-59-2010}
\bibinfo{author}{\bibfnamefont{R.}~\bibnamefont{van Roij}},
  \bibinfo{author}{\bibfnamefont{M.}~\bibnamefont{Dijkstra}}, \bibnamefont{and}
  \bibinfo{author}{\bibfnamefont{J.-P.} \bibnamefont{Hansen}},
  \bibinfo{journal}{Phys. Rev. E} \textbf{\bibinfo{volume}{59}},
  \bibinfo{pages}{2010} (\bibinfo{year}{1999}).

\bibitem[{\citenamefont{Donnan}(1924)}]{donnan1924}
\bibinfo{author}{\bibfnamefont{F.~G.} \bibnamefont{Donnan}},
  \bibinfo{journal}{Chem. Rev.} \textbf{\bibinfo{volume}{1}},
  \bibinfo{pages}{73} (\bibinfo{year}{1924}).

\bibitem[{\citenamefont{Overbeek}(1956)}]{overbeek1956}
\bibinfo{author}{\bibfnamefont{J.~T.~G.} \bibnamefont{Overbeek}},
  \bibinfo{journal}{Prog. Biophys. Biophys. Chem.}
  \textbf{\bibinfo{volume}{6}}, \bibinfo{pages}{57} (\bibinfo{year}{1956}).

\bibitem[{\citenamefont{Alexander et~al.}(1984)\citenamefont{Alexander,
  Chaikin, Grant, Morales, Pincus, and Hone}}]{JCP-80-5776}
\bibinfo{author}{\bibfnamefont{S.}~\bibnamefont{Alexander}},
  \bibinfo{author}{\bibfnamefont{P.~M.} \bibnamefont{Chaikin}},
  \bibinfo{author}{\bibfnamefont{P.}~\bibnamefont{Grant}},
  \bibinfo{author}{\bibfnamefont{G.}~\bibnamefont{Morales}},
  \bibinfo{author}{\bibfnamefont{P.}~\bibnamefont{Pincus}}, \bibnamefont{and}
  \bibinfo{author}{\bibfnamefont{D.}~\bibnamefont{Hone}}, \bibinfo{journal}{J.
  Chem. Phys.} \textbf{\bibinfo{volume}{80}}, \bibinfo{pages}{5776}
  (\bibinfo{year}{1984}).

\bibitem[{\citenamefont{Trizac and Levin}(2004)}]{PRE-69-031403}
\bibinfo{author}{\bibfnamefont{E.}~\bibnamefont{Trizac}} \bibnamefont{and}
  \bibinfo{author}{\bibfnamefont{Y.}~\bibnamefont{Levin}},
  \bibinfo{journal}{Phys. Rev. E} \textbf{\bibinfo{volume}{69}},
  \bibinfo{pages}{031403} (\bibinfo{year}{2004}).

\bibitem[{\citenamefont{Levin et~al.}(2003)\citenamefont{Levin, Trizac, and
  Bocquet}}]{JPCM-15-S3523}
\bibinfo{author}{\bibfnamefont{Y.}~\bibnamefont{Levin}},
  \bibinfo{author}{\bibfnamefont{E.}~\bibnamefont{Trizac}}, \bibnamefont{and}
  \bibinfo{author}{\bibfnamefont{L.}~\bibnamefont{Bocquet}},
  \bibinfo{journal}{J. Phys.: Condens. Matter} \textbf{\bibinfo{volume}{15}},
  \bibinfo{pages}{S3523} (\bibinfo{year}{2003}).

\bibitem[{\citenamefont{Trizac et~al.}(2003)\citenamefont{Trizac, Bocquet,
  Aubouy, and von Gr{\"u}nberg}}]{Lang-19-4027}
\bibinfo{author}{\bibfnamefont{E.}~\bibnamefont{Trizac}},
  \bibinfo{author}{\bibfnamefont{L.}~\bibnamefont{Bocquet}},
  \bibinfo{author}{\bibfnamefont{M.}~\bibnamefont{Aubouy}}, \bibnamefont{and}
  \bibinfo{author}{\bibfnamefont{H.}~\bibnamefont{von Gr{\"u}nberg}},
  \bibinfo{journal}{Langmuir} \textbf{\bibinfo{volume}{19}},
  \bibinfo{pages}{4027} (\bibinfo{year}{2003}).

\bibitem[{\citenamefont{Belloni}(1998)}]{CollSutfA-140-227}
\bibinfo{author}{\bibfnamefont{L.}~\bibnamefont{Belloni}},
  \bibinfo{journal}{Colloids Surf. A} \textbf{\bibinfo{volume}{140}},
  \bibinfo{pages}{227} (\bibinfo{year}{1998}).

\bibitem[{\citenamefont{Bocquet et~al.}(2002)\citenamefont{Bocquet, Trizac, and
  Aubouy}}]{JCP-117-8138}
\bibinfo{author}{\bibfnamefont{L.}~\bibnamefont{Bocquet}},
  \bibinfo{author}{\bibfnamefont{E.}~\bibnamefont{Trizac}}, \bibnamefont{and}
  \bibinfo{author}{\bibfnamefont{M.}~\bibnamefont{Aubouy}},
  \bibinfo{journal}{J. Chem. Phys.} \textbf{\bibinfo{volume}{117}},
  \bibinfo{pages}{8138} (\bibinfo{year}{2002}).

\bibitem[{\citenamefont{Levin et~al.}(1998)\citenamefont{Levin, Barbosa, and
  Tamashiro}}]{EPL-41-123}
\bibinfo{author}{\bibfnamefont{Y.}~\bibnamefont{Levin}},
  \bibinfo{author}{\bibfnamefont{M.~C.} \bibnamefont{Barbosa}},
  \bibnamefont{and} \bibinfo{author}{\bibfnamefont{M.~N.}
  \bibnamefont{Tamashiro}}, \bibinfo{journal}{Europhys. Lett.}
  \textbf{\bibinfo{volume}{41}}, \bibinfo{pages}{123} (\bibinfo{year}{1998}).

\bibitem[{\citenamefont{Tamashiro and Schiessel}(2003)}]{JCP-119-1855}
\bibinfo{author}{\bibfnamefont{M.~N.} \bibnamefont{Tamashiro}}
  \bibnamefont{and}
  \bibinfo{author}{\bibfnamefont{H.}~\bibnamefont{Schiessel}},
  \bibinfo{journal}{J. Chem. Phys.} \textbf{\bibinfo{volume}{119}},
  \bibinfo{pages}{1855} (\bibinfo{year}{2003}).

\bibitem[{\citenamefont{von Gr{\"u}nberg et~al.}(2001)\citenamefont{von
  Gr{\"u}nberg, van Roij, and Klein}}]{EPL-55-580}
\bibinfo{author}{\bibfnamefont{H.~H.} \bibnamefont{von Gr{\"u}nberg}},
  \bibinfo{author}{\bibfnamefont{R.}~\bibnamefont{van Roij}}, \bibnamefont{and}
  \bibinfo{author}{\bibfnamefont{G.}~\bibnamefont{Klein}},
  \bibinfo{journal}{Europhys. Lett.} \textbf{\bibinfo{volume}{55}},
  \bibinfo{pages}{580} (\bibinfo{year}{2001}).

\bibitem[{\citenamefont{Zoetekouw and van Roij}({\natexlab{a}})}]{bas_multcel}
\bibinfo{author}{\bibfnamefont{B.}~\bibnamefont{Zoetekouw}} \bibnamefont{and}
  \bibinfo{author}{\bibfnamefont{R.}~\bibnamefont{van Roij}},
  \emph{\bibinfo{note}{unpublished}}.

\bibitem[{\citenamefont{Evans}(1979)}]{AdvPhys-28-143}
\bibinfo{author}{\bibfnamefont{R.}~\bibnamefont{Evans}}, \bibinfo{journal}{Adv.
  Phys.} \textbf{\bibinfo{volume}{28}}, \bibinfo{pages}{143}
  (\bibinfo{year}{1979}).

\bibitem[{\citenamefont{Evans}(1992)}]{Evans1992}
\bibinfo{author}{\bibfnamefont{R.}~\bibnamefont{Evans}},
  \emph{\bibinfo{title}{Fundamentals of inhomogeneous fluids}}
  (\bibinfo{publisher}{Dekker, New York}, \bibinfo{year}{1992}), chap.
  \bibinfo{chapter}{Density functionals in the theory of nonuniform fluids},
  pp. \bibinfo{pages}{85--175}.

\bibitem[{\citenamefont{L{\"o}wen}(2002)}]{JPCM-14-11897}
\bibinfo{author}{\bibfnamefont{H.}~\bibnamefont{L{\"o}wen}},
  \bibinfo{journal}{J. Phys.: Condens. Matter} \textbf{\bibinfo{volume}{14}},
  \bibinfo{pages}{11897} (\bibinfo{year}{2002}).

\bibitem[{\citenamefont{Marcus}(1955)}]{marcus1955}
\bibinfo{author}{\bibfnamefont{R.~A.} \bibnamefont{Marcus}},
  \bibinfo{journal}{J. Chem. Phys.} \textbf{\bibinfo{volume}{23}},
  \bibinfo{pages}{1057} (\bibinfo{year}{1955}).

\bibitem[{\citenamefont{Fushiki}(1992)}]{fushiki_JCP_97}
\bibinfo{author}{\bibfnamefont{M.}~\bibnamefont{Fushiki}}, \bibinfo{journal}{J.
  Chem. Phys.} \textbf{\bibinfo{volume}{97}}, \bibinfo{pages}{6700}
  (\bibinfo{year}{1992}).

\bibitem[{\citenamefont{L{\"o}wen et~al.}(1992)\citenamefont{L{\"o}wen, Madden,
  and Hansen}}]{PRL-68-1081}
\bibinfo{author}{\bibfnamefont{H.}~\bibnamefont{L{\"o}wen}},
  \bibinfo{author}{\bibfnamefont{P.~A.} \bibnamefont{Madden}},
  \bibnamefont{and} \bibinfo{author}{\bibfnamefont{J.~P.}
  \bibnamefont{Hansen}}, \bibinfo{journal}{Phys. Rev. Lett.}
  \textbf{\bibinfo{volume}{68}}, \bibinfo{pages}{1081} (\bibinfo{year}{1992}).

\bibitem[{\citenamefont{L{\"o}wen et~al.}(1993)\citenamefont{L{\"o}wen, Hansen,
  and Madden}}]{JCP-92-3275}
\bibinfo{author}{\bibfnamefont{H.}~\bibnamefont{L{\"o}wen}},
  \bibinfo{author}{\bibfnamefont{J.-P.} \bibnamefont{Hansen}},
  \bibnamefont{and} \bibinfo{author}{\bibfnamefont{P.~A.}
  \bibnamefont{Madden}}, \bibinfo{journal}{J. Chem. Phys.}
  \textbf{\bibinfo{volume}{92}}, \bibinfo{pages}{3275} (\bibinfo{year}{1993}).

\bibitem[{\citenamefont{Dobnikar
  et~al.}(2003{\natexlab{a}})\citenamefont{Dobnikar, Chen, Rzehak, and von
  Gr{\"u}nberg}}]{dobnikar2}
\bibinfo{author}{\bibfnamefont{J.}~\bibnamefont{Dobnikar}},
  \bibinfo{author}{\bibfnamefont{Y.}~\bibnamefont{Chen}},
  \bibinfo{author}{\bibfnamefont{R.}~\bibnamefont{Rzehak}}, \bibnamefont{and}
  \bibinfo{author}{\bibfnamefont{H.~H.} \bibnamefont{von Gr{\"u}nberg}},
  \bibinfo{journal}{J. Phys.: Condens. Matter} \textbf{\bibinfo{volume}{15}},
  \bibinfo{pages}{S263} (\bibinfo{year}{2003}{\natexlab{a}}).

\bibitem[{\citenamefont{Dobnikar
  et~al.}(2003{\natexlab{b}})\citenamefont{Dobnikar, Chen, Rzehak, and von
  Gr{\"u}nberg}}]{dobnikar1}
\bibinfo{author}{\bibfnamefont{J.}~\bibnamefont{Dobnikar}},
  \bibinfo{author}{\bibfnamefont{Y.}~\bibnamefont{Chen}},
  \bibinfo{author}{\bibfnamefont{R.}~\bibnamefont{Rzehak}}, \bibnamefont{and}
  \bibinfo{author}{\bibfnamefont{H.~H.} \bibnamefont{von Gr{\"u}nberg}},
  \bibinfo{journal}{J. Chem. Phys.} \textbf{\bibinfo{volume}{119}},
  \bibinfo{pages}{4971} (\bibinfo{year}{2003}{\natexlab{b}}).

\bibitem[{\citenamefont{Dobnikar
  et~al.}(2003{\natexlab{c}})\citenamefont{Dobnikar, Rzehak, and von
  Gr{\"u}nberg}}]{dobnikar3}
\bibinfo{author}{\bibfnamefont{J.}~\bibnamefont{Dobnikar}},
  \bibinfo{author}{\bibfnamefont{R.}~\bibnamefont{Rzehak}}, \bibnamefont{and}
  \bibinfo{author}{\bibfnamefont{H.~H.} \bibnamefont{von Gr{\"u}nberg}},
  \bibinfo{journal}{Europhys. Lett.} \textbf{\bibinfo{volume}{61}},
  \bibinfo{pages}{695} (\bibinfo{year}{2003}{\natexlab{c}}).

\bibitem[{\citenamefont{Deserno and von Gr{\"u}nberg}(2002)}]{PRE-66-011401}
\bibinfo{author}{\bibfnamefont{M.}~\bibnamefont{Deserno}} \bibnamefont{and}
  \bibinfo{author}{\bibfnamefont{H.-H.} \bibnamefont{von Gr{\"u}nberg}},
  \bibinfo{journal}{Phys. Rev. E} \textbf{\bibinfo{volume}{66}},
  \bibinfo{pages}{011401} (\bibinfo{year}{2002}).

\bibitem[{\citenamefont{Deserno and Holm}(2002)}]{deserno-holm-2001}
\bibinfo{author}{\bibfnamefont{M.}~\bibnamefont{Deserno}} \bibnamefont{and}
  \bibinfo{author}{\bibfnamefont{C.}~\bibnamefont{Holm}},
  \emph{\bibinfo{title}{Electrostatic Effects in Soft Matter and Biophysics}}
  (\bibinfo{publisher}{Kluwer, Dordrecht}, \bibinfo{year}{2002}),
  vol.~\bibinfo{volume}{46} of \emph{\bibinfo{series}{NATO Science Series II
  --- Mathematics, Physics and Chemistry}}, chap. \bibinfo{chapter}{Cell model
  and Poisson--Boltzmann theory: a brief introduction}, ISBN
  \bibinfo{isbn}{1-4020-0197-5}.

\bibitem[{\citenamefont{Ra{\c{s}}a et~al.}(2005)\citenamefont{Ra{\c{s}}a,
  Ern{\'e}, Zoetekouw, van Roij, and Philipse}}]{mircea}
\bibinfo{author}{\bibfnamefont{M.}~\bibnamefont{Ra{\c{s}}a}},
  \bibinfo{author}{\bibfnamefont{B.~H.} \bibnamefont{Ern{\'e}}},
  \bibinfo{author}{\bibfnamefont{B.}~\bibnamefont{Zoetekouw}},
  \bibinfo{author}{\bibfnamefont{R.}~\bibnamefont{van Roij}}, \bibnamefont{and}
  \bibinfo{author}{\bibfnamefont{A.~P.} \bibnamefont{Philipse}},
  \bibinfo{journal}{J. Phys.: Condens. Matter} \textbf{\bibinfo{volume}{17}},
  \bibinfo{pages}{2293} (\bibinfo{year}{2005}).

\bibitem[{\citenamefont{Ra{\c{s}}a and Philipse}(2004)}]{mircea_nature}
\bibinfo{author}{\bibfnamefont{M.}~\bibnamefont{Ra{\c{s}}a}} \bibnamefont{and}
  \bibinfo{author}{\bibfnamefont{A.~P.} \bibnamefont{Philipse}},
  \bibinfo{journal}{Nature} \textbf{\bibinfo{volume}{429}},
  \bibinfo{pages}{857} (\bibinfo{year}{2004}).

\bibitem[{\citenamefont{Hansen and McDonald}(1986)}]{HMcD}
\bibinfo{author}{\bibfnamefont{J.-P.} \bibnamefont{Hansen}} \bibnamefont{and}
  \bibinfo{author}{\bibfnamefont{I.~R.} \bibnamefont{McDonald}},
  \emph{\bibinfo{title}{Theory of simple liquids}}
  (\bibinfo{publisher}{Academic Press, London}, \bibinfo{year}{1986}).

\bibitem[{\citenamefont{Isihara}(1968)}]{Isihara}
\bibinfo{author}{\bibfnamefont{A.}~\bibnamefont{Isihara}}, \bibinfo{journal}{J.
  Phys. A (Proc. Phys. Soc.)} \textbf{\bibinfo{volume}{1}},
  \bibinfo{pages}{539} (\bibinfo{year}{1968}).

\bibitem[{\citenamefont{Gibbs}(1902)}]{Gibbs}
\bibinfo{author}{\bibfnamefont{J.~W.} \bibnamefont{Gibbs}},
  \emph{\bibinfo{title}{Elementary principles in Statistical Mechanics}}
  (\bibinfo{publisher}{Oxford University Press, Oxford}, \bibinfo{year}{1902}),
  \emph{\bibinfo{note}{chapter XI}}.

\bibitem[{\citenamefont{Bogoliubov}(1954)}]{Bogoliubov}
\bibinfo{author}{\bibfnamefont{N.~N.} \bibnamefont{Bogoliubov}},
  \bibinfo{journal}{Dokl. Akad. Nauk SSSR} \textbf{\bibinfo{volume}{119}},
  \bibinfo{pages}{244} (\bibinfo{year}{1954}).

\bibitem[{\citenamefont{Carnahan and Starling}(1969)}]{JCP-51-635}
\bibinfo{author}{\bibfnamefont{N.~F.} \bibnamefont{Carnahan}} \bibnamefont{and}
  \bibinfo{author}{\bibfnamefont{K.~E.} \bibnamefont{Starling}},
  \bibinfo{journal}{J. Chem. Phys.} \textbf{\bibinfo{volume}{51}},
  \bibinfo{pages}{635} (\bibinfo{year}{1969}).

\bibitem[{\citenamefont{Verlet and Weis}(1972)}]{PRA-5-939}
\bibinfo{author}{\bibfnamefont{L.}~\bibnamefont{Verlet}} \bibnamefont{and}
  \bibinfo{author}{\bibfnamefont{J.-J.} \bibnamefont{Weis}},
  \bibinfo{journal}{Phys. Rev. A} \textbf{\bibinfo{volume}{5}},
  \bibinfo{pages}{939} (\bibinfo{year}{1972}).

\bibitem[{\citenamefont{Bravo~Yuste and Santos}(1991)}]{PRA-43-5418}
\bibinfo{author}{\bibfnamefont{S.}~\bibnamefont{Bravo~Yuste}} \bibnamefont{and}
  \bibinfo{author}{\bibfnamefont{A.}~\bibnamefont{Santos}},
  \bibinfo{journal}{Phys. Rev. A} \textbf{\bibinfo{volume}{43}},
  \bibinfo{pages}{5418} (\bibinfo{year}{1991}).

\bibitem[{\citenamefont{Bravo~Yuste et~al.}(1996)\citenamefont{Bravo~Yuste,
  {L{\'o}pez de Haro}, and Santos}}]{PRE-53-4820}
\bibinfo{author}{\bibfnamefont{S.}~\bibnamefont{Bravo~Yuste}},
  \bibinfo{author}{\bibfnamefont{M.}~\bibnamefont{{L{\'o}pez de Haro}}},
  \bibnamefont{and} \bibinfo{author}{\bibfnamefont{A.}~\bibnamefont{Santos}},
  \bibinfo{journal}{Phys. Rev. E} \textbf{\bibinfo{volume}{53}},
  \bibinfo{pages}{4820} (\bibinfo{year}{1996}).

\bibitem[{\citenamefont{Ziman}(1972)}]{ziman}
\bibinfo{author}{\bibfnamefont{J.~M.} \bibnamefont{Ziman}},
  \emph{\bibinfo{title}{Theory of Solids}} (\bibinfo{publisher}{Cambridge
  University Press, Cambridge}, \bibinfo{year}{1972}), \bibinfo{edition}{2nd}
  ed.

\bibitem[{\citenamefont{Ashcroft and Mermin}(1976)}]{ashcroft}
\bibinfo{author}{\bibfnamefont{N.~W.} \bibnamefont{Ashcroft}} \bibnamefont{and}
  \bibinfo{author}{\bibfnamefont{N.~D.} \bibnamefont{Mermin}},
  \emph{\bibinfo{title}{Solid state physics}} (\bibinfo{publisher}{Saunders
  College Publishing}, \bibinfo{year}{1976}).

\bibitem[{\citenamefont{Shih et~al.}(1987)\citenamefont{Shih, Aksay, and
  Kikuchi}}]{JCP-86-5127}
\bibinfo{author}{\bibfnamefont{W.~Y.} \bibnamefont{Shih}},
  \bibinfo{author}{\bibfnamefont{I.~A.} \bibnamefont{Aksay}}, \bibnamefont{and}
  \bibinfo{author}{\bibfnamefont{R.}~\bibnamefont{Kikuchi}},
  \bibinfo{journal}{J. Chem. Phys.} \textbf{\bibinfo{volume}{86}},
  \bibinfo{pages}{5127} (\bibinfo{year}{1987}).

\bibitem[{\citenamefont{Levin}(2000)}]{JCP-113-9722}
\bibinfo{author}{\bibfnamefont{Y.}~\bibnamefont{Levin}}, \bibinfo{journal}{J.
  Chem. Phys.} \textbf{\bibinfo{volume}{113}}, \bibinfo{pages}{9722}
  (\bibinfo{year}{2000}).

\bibitem[{\citenamefont{Hynninen et~al.}(2005)\citenamefont{Hynninen, Dijkstra,
  and Panagiotopoulos}}]{antti-preprint}
\bibinfo{author}{\bibfnamefont{A.-P.} \bibnamefont{Hynninen}},
  \bibinfo{author}{\bibfnamefont{M.}~\bibnamefont{Dijkstra}}, \bibnamefont{and}
  \bibinfo{author}{\bibfnamefont{A.~Z.} \bibnamefont{Panagiotopoulos}},
  \bibinfo{journal}{J. Chem. Phys.} \textbf{\bibinfo{volume}{123}},
  \bibinfo{pages}{84903} (\bibinfo{year}{2005}).

\bibitem[{\citenamefont{Fisher and Levin}(1993)}]{PRL-71-3826}
\bibinfo{author}{\bibfnamefont{M.~E.} \bibnamefont{Fisher}} \bibnamefont{and}
  \bibinfo{author}{\bibfnamefont{Y.}~\bibnamefont{Levin}},
  \bibinfo{journal}{Phys. Rev. Lett.} \textbf{\bibinfo{volume}{71}},
  \bibinfo{pages}{3826} (\bibinfo{year}{1993}).

\bibitem[{\citenamefont{Caillol et~al.}(1997)\citenamefont{Caillol, Levesque,
  and Weis}}]{JCP-107-1565}
\bibinfo{author}{\bibfnamefont{J.~M.} \bibnamefont{Caillol}},
  \bibinfo{author}{\bibfnamefont{D.}~\bibnamefont{Levesque}}, \bibnamefont{and}
  \bibinfo{author}{\bibfnamefont{J.~J.} \bibnamefont{Weis}},
  \bibinfo{journal}{J. Chem. Phys.} \textbf{\bibinfo{volume}{107}},
  \bibinfo{pages}{1565} (\bibinfo{year}{1997}).

\bibitem[{\citenamefont{Orkoulas and Panagiotopoulos}(1999)}]{JCP-110-1581}
\bibinfo{author}{\bibfnamefont{G.}~\bibnamefont{Orkoulas}} \bibnamefont{and}
  \bibinfo{author}{\bibfnamefont{A.~Z.} \bibnamefont{Panagiotopoulos}},
  \bibinfo{journal}{J. Chem. Phys.} \textbf{\bibinfo{volume}{110}},
  \bibinfo{pages}{1581} (\bibinfo{year}{1999}).

\bibitem[{\citenamefont{Yan and de~Pablo}(1999)}]{JCP-111-9509}
\bibinfo{author}{\bibfnamefont{Q.}~\bibnamefont{Yan}} \bibnamefont{and}
  \bibinfo{author}{\bibfnamefont{J.~J.} \bibnamefont{de~Pablo}},
  \bibinfo{journal}{J. Chem. Phys.} \textbf{\bibinfo{volume}{111}},
  \bibinfo{pages}{9509} (\bibinfo{year}{1999}).

\bibitem[{\citenamefont{Levin and Fisher}(1995)}]{levin_fisher}
\bibinfo{author}{\bibfnamefont{Y.}~\bibnamefont{Levin}} \bibnamefont{and}
  \bibinfo{author}{\bibfnamefont{M.~E.} \bibnamefont{Fisher}},
  \bibinfo{journal}{Physica A} \textbf{\bibinfo{volume}{225}},
  \bibinfo{pages}{164} (\bibinfo{year}{1995}).

\bibitem[{\citenamefont{Panagiotopoulos and Fisher}(2002)}]{PRL-88-045701}
\bibinfo{author}{\bibfnamefont{A.~Z.} \bibnamefont{Panagiotopoulos}}
  \bibnamefont{and} \bibinfo{author}{\bibfnamefont{M.~E.}
  \bibnamefont{Fisher}}, \bibinfo{journal}{Phys. Rev. Lett.}
  \textbf{\bibinfo{volume}{88}}, \bibinfo{pages}{045701}
  (\bibinfo{year}{2002}).

\bibitem[{\citenamefont{Biesheuvel}(2004)}]{JPCM-16-L499}
\bibinfo{author}{\bibfnamefont{P.~M.} \bibnamefont{Biesheuvel}},
  \bibinfo{journal}{J. Phys.: Condens. Matter} \textbf{\bibinfo{volume}{16}},
  \bibinfo{pages}{L499} (\bibinfo{year}{2004}).

\bibitem[{\citenamefont{van Roij}(2003)}]{JPCM-15-S3569}
\bibinfo{author}{\bibfnamefont{R.}~\bibnamefont{van Roij}},
  \bibinfo{journal}{J. Phys.: Condens. Matter} \textbf{\bibinfo{volume}{15}},
  \bibinfo{pages}{S3569} (\bibinfo{year}{2003}).

\bibitem[{\citenamefont{Linse}(2000)}]{JCP-113-4359}
\bibinfo{author}{\bibfnamefont{P.}~\bibnamefont{Linse}}, \bibinfo{journal}{J.
  Chem. Phys.} \textbf{\bibinfo{volume}{113}}, \bibinfo{pages}{4359}
  (\bibinfo{year}{2000}).

\bibitem[{\citenamefont{Zoetekouw and van
  Roij}({\natexlab{b}})}]{bas_metaphase}
\bibinfo{author}{\bibfnamefont{B.}~\bibnamefont{Zoetekouw}} \bibnamefont{and}
  \bibinfo{author}{\bibfnamefont{R.}~\bibnamefont{van Roij}},
  \emph{\bibinfo{note}{in preparation}}.

\bibitem[{\citenamefont{van Roij and Evans}(1999)}]{JPCM-11-10047}
\bibinfo{author}{\bibfnamefont{R.}~\bibnamefont{van Roij}} \bibnamefont{and}
  \bibinfo{author}{\bibfnamefont{R.}~\bibnamefont{Evans}}, \bibinfo{journal}{J.
  Phys.: Condens. Matter} \textbf{\bibinfo{volume}{11}}, \bibinfo{pages}{10047}
  (\bibinfo{year}{1999}).

\bibitem[{\citenamefont{van Roij}(2000)}]{JPCM-12-A263}
\bibinfo{author}{\bibfnamefont{R.}~\bibnamefont{van Roij}},
  \bibinfo{journal}{J. Phys.: Condens. Matter} \textbf{\bibinfo{volume}{12}},
  \bibinfo{pages}{A263} (\bibinfo{year}{2000}).

\bibitem[{\citenamefont{Fisher}(1996)}]{JPCM-8-9103}
\bibinfo{author}{\bibfnamefont{M.~E.} \bibnamefont{Fisher}},
  \bibinfo{journal}{J. Phys.: Condens. Matter} \textbf{\bibinfo{volume}{8}},
  \bibinfo{pages}{9103} (\bibinfo{year}{1996}).

\bibitem[{\citenamefont{Monovoukas and Gast}(1989)}]{JCI-128-533}
\bibinfo{author}{\bibfnamefont{Y.}~\bibnamefont{Monovoukas}} \bibnamefont{and}
  \bibinfo{author}{\bibfnamefont{A.~P.} \bibnamefont{Gast}},
  \bibinfo{journal}{J. Colloid Interface Sci.} \textbf{\bibinfo{volume}{128}},
  \bibinfo{pages}{533} (\bibinfo{year}{1989}).

\end{thebibliography}

\end{document}